\documentclass{emulateapj}
\slugcomment{Accepted for publication in The Astrophysical Journal}

\usepackage{apjfonts}
\usepackage{mathptmx}
\usepackage{natbib}
\usepackage{amsmath}
\usepackage{hyphenat}
\newcommand{\height}{3.4in}
\newcommand{\heightb}{5.9in}
\newcommand{\beq}{\begin{equation}}
\newcommand{\eeq}{\end{equation}}
\begin{document}

\title{Reconstructing Redshift Distributions with Cross-Correlations: Tests and an Optimized Recipe}
\author{Daniel J. Matthews and Jeffrey A. Newman}
\affil{Department of Physics and Astronomy, University of Pittsburgh, 3941 O'Hara Street, Pittsburgh, PA 15260}
\email{djm70@pitt.edu, janewman@pitt.edu}

\begin{abstract}

Many of the cosmological tests to be performed by planned dark energy experiments will require extremely well-characterized photometric redshift measurements. Current estimates for cosmic shear are that the true mean redshift of the objects in each photo-z bin must be known to better than $0.002(1+z)$, and the width of the bin must be known to $\sim0.003(1+z)$ if errors in cosmological measurements are not to be degraded significantly.  A conventional approach is to calibrate these photometric redshifts with large sets of spectroscopic redshifts. However, at the depths probed by Stage III surveys (such as DES), let alone Stage IV (LSST, JDEM, Euclid), existing large redshift samples have all been highly (25-60\%) incomplete, with a strong dependence of success rate on both redshift and galaxy properties. A powerful alternative approach is to exploit the clustering of galaxies to perform photometric redshift calibrations. Measuring the two-point angular cross\hyp{}correlation between objects in some photometric redshift bin and objects with known spectroscopic redshift, as a function of the spectroscopic z, allows the true redshift distribution of a photometric sample to be reconstructed in detail, even if it includes objects too faint for spectroscopy or if spectroscopic samples are highly incomplete. We test this technique using mock DEEP2 Galaxy Redshift survey light cones constructed from the Millennium Simulation semi-analytic galaxy catalogs. From this realistic test, which incorporates the effects of galaxy bias evolution and cosmic variance, we find that the true redshift distribution of a photometric sample can, in fact, be determined accurately with cross\hyp{}correlation techniques.  We also compare the empirical error in the reconstruction of redshift distributions to previous analytic predictions, finding that additional components must be included in error budgets to match the simulation results.  This extra error contribution is small for surveys which sample large areas of sky ($> \sim $10-100 degrees), but dominant for $\sim 1$ square degree fields. We conclude by presenting a step-by-step, optimized recipe for reconstructing redshift distributions from cross\hyp{}correlation information using standard correlation measurements.

\end{abstract}


\section{Introduction}
\label{sec:intro}
For many years it was thought that the expansion of the universe should be slowing due to the gravitational attraction of matter, but measurements of Type Ia supernovae and other observations have shown that the expansion rate is in fact accelerating \citep{1998AJ....116.1009R,1999ApJ...517..565P}. This accelerating expansion is generally attributed to an unknown component of the energy density of the universe commonly referred to as ``dark energy.'' One of the goals of future cosmological probes (e.g. LSST, JDEM, and Euclid) \citep{2001ASPC..232..347T, 2005ASPC..339...95T,2009arXiv0901.0721A, 2010arXiv1001.3349B}, is to determine constraints on dark energy equation of state parameters, e.g. $w \equiv P/\rho$ and $\mathrm{w_a} \equiv dw/da$ \citep{2007IJMPD..16.1581J}, where $P$ is the pressure from dark energy, $\rho$ is its mass density, and $a$ is the scale factor of the universe (normalized to be 1 today). 

In order for these experiments to be successful, we require information about the redshift of all objects used to make measurements. However, it is impractical to measure spectroscopic redshifts for hundreds of millions of galaxies, especially extremely faint ones. We can measure the redshift of many more objects from photometric information, e.g. by using a large set of spectroscopic redshifts to create templates of how color varies with redshift \citep{1995AJ....110.2655C}. However current and future spectroscopic surveys will be highly incomplete due to selection biases dependent on redshift and galaxy properties \citep{2006MNRAS.370..198C}. Because of this, along with the catastrophic photometric errors\footnote{such as contamination from overlapping or unresolved objects; this is a frequent problem in deep surveys, particularly at high redshifts, cf. Newman et al. 2010} that can occur at a significant ($\sim 1\%$) rate \citep{2009ApJ...699..958S, 2010MNRAS.401.1399B}, photometric redshifts are not as well understood as redshifts determined spectroscopically. If future dark energy experiments are to reach their goals, it is necessary to develop a method of calibrating photometric redshifts with high precision \citep{2006astro.ph..9591A, 2006MNRAS.366..101H, 2006ApJ...636...21M}. Current projections for LSST cosmic shear measurements estimate that the true mean redshift of objects in each photo-z bin must be known to better than $\sim0.002(1+z)$ \citep{2006ApJ...644..663Z, 2006JCAP...08..008Z, 2006ApJ...652..857K, 2006AIPC..870...44T} with stringent requirements on the fraction of unconstrained catastrophic outliers \citep{2010arXiv1002.3383H}, while the width of the bin must be known to $\sim0.003(1+z)$ \citep{2009arXiv0912.0201L}.

In this paper we test a new technique for calibrating photometric redshifts measured by other algorithms, which exploits the fact that objects at similar redshifts tend to cluster with each other. If we have two galaxy samples, one with only photometric information and the other consisting of objects with known spectroscopic redshifts, we can measure the angular cross\hyp{}correlation between objects in the photometric sample and the spectroscopic sample as a function of spectroscopic $z$. This clustering will depend on both the intrinsic clustering of the samples with each other and the degree to which the samples overlap in redshift. Autocorrelation measurements for each sample give information about their intrinsic clustering, which can be used to break the degeneracy between these two contributions. The principal advantage of this technique is that, while the two sets of objects should overlap in redshift and on the sky, it is not necessary for the spectroscopic sample to be complete at any given redshift. Therefore it is possible to use only the brightest objects at a given $z$, from which it is much easier to obtain secure redshift measurements, to calibrate photometric redshifts.  Even systematic incompleteness (e.g. failing to obtain redshifts for galaxies of specific types) in the spectroscopic sample is not a problem, so long as the full redshift range is sampled. This method is effective even when the two samples do not have similar properties (e.g. differing luminosity and bias).

We here describe a complete end-to-end implementation of cross\hyp{}correlation methods for calibrating photometric redshifts and present the results of applying these algorithms to realistic mock catalogs. Throughout the paper we assume a flat $\Lambda$CDM cosmology with $\Omega_m$=0.3, $\Omega_{\Lambda}$=0.7, and Hubble parameter $H_0=100h$ km s$^{-1}$ Mpc$^{-1}$, where we have assumed $h$=0.72, matching the Millennium simulations, where it is not explicitly included in formulae. In \S \ref{sec:datasets} we describe the catalog and data sets used to test cross\hyp{}correlation methods. In \S \ref{sec:method} we provide a description of the reconstruction techniques used in detail, and in \S \ref{sec:results} we provide the results of the calculation. In \S \ref{sec:conclusion} we conclude, as well as give a more concise description of the steps taken, providing a recipe for cross\hyp{}correlation photometric redshift calibration.

\section{Data Sets}
\label{sec:datasets}
To test this method, it is necessary to construct two samples of galaxies, one with known redshift (``spectroscopic'') and the other unknown (``photometric''). We have done this using mock DEEP2 Redshift Survey light cones produced by Darren Croton. A total of 24 light cones were constructed by taking lines-of-sight through the Millennium Simulation halo catalog \citep{2006astro.ph..8019L} with the redshift of the simulation cube used increasing with distance from the observer \citep{2007MNRAS.376....2K}. The light cones were then populated with galaxies using a semi-analytic model whose parameters were chosen to reproduce local galaxy properties \citep{2006MNRAS.365...11C}. Each light cone covers the range $0.10<z<1.5$ and corresponds to a $0.5\times2.0$ degree region of sky. The galaxies in this mock catalog will have properties (including color, luminosity, and large-scale structure bias) which vary with redshift due to the same factors believed to affect real galaxy evolution. The semi-analytic model used is certainly imperfect, but yields samples of galaxies that pose the same difficulties (e.g. bias evolution and differences in clustering between bright and faint objects) as real surveys will exhibit; they therefore provide a realistic test of our ability to reconstruct redshift distributions of faint samples using spectroscopy of only a brighter subset.

The spectroscopic sample is generated by selecting 60\% of objects with observed $R$-band magnitude $R<24.1$, which gives a sample whose characteristics resemble the DEEP2 Galaxy Redshift survey (Newman et al. 2010, in prep.). The mean number of spectroscopic objects over the 24 light cones is $35,574$. The size of this sample is comparable to the number of objects predicted to be needed for calibration using template-based methods ($\sim10^5$ \citep{2009arXiv0912.0201L, 2008ApJ...682...39M}). However, this sample differs greatly in what it contains: it consists only of relatively bright objects, rather than having to be a statistically complete sample extending as faint as the objects to which photometric redshifts will be applied (a necessity for accurate training or template development, as the spectral energy distributions of faint galaxies are observed to lie outside the range luminous galaxies cover, both at $z\sim 0$ and $z\sim 1$ \citep{2006ApJ...647..853W, 2010PASP..122..485M}.  Studies such as \citet{2010MNRAS.401.1399B} have assumed for such projections that 99.9\% redshift success can be achieved for faint galaxy samples (e.g. of photometric-redshift outliers); however, that is a failure rate more than two orders of magnitude lower than that actually achieved by current large surveys on 10-meter class telescopes such as VVDS \citep{2005A&A...439..845L}, ZCOSMOS \citep{2007ApJS..172...70L}, or DEEP2 (Newman et al. 2010, in prep.), surveys which are 1.5-5 magnitudes shallower than the limits of Stage III and Stage IV surveys such as DES and LSST.  In contrast, as noted in \S \ref{sec:intro}, the cross\hyp{}correlation techniques we focus on in this paper do not require a complete spectroscopic sample, and hence do not require improvements in redshift success over existing projects to provide an accurate calibration.

The other sample, referred to hereafter as the photometric sample, is constructed by selecting objects in the mock catalog down to the faintest magnitudes available, with the probability of inclusion a Gaussian with $\langle z \rangle = 0.75$ and $\sigma_z = 0.20$. This emulates choosing a set of objects which have been placed in a single photometric redshift bin by some algorithm with Gaussian errors.  It should be noted that, since the redshift distribution of the mock catalog we select from is not uniform, the resulting redshift distribution of the photometric sample is not a pure Gaussian. The overall redshift distribution of all objects in the catalog is fit well using a 5th degree polynomial, so the net distribution of the photometric sample can be well represented by the product of this polynomial and a Gaussian. After applying this Gaussian selection to the mock catalog, we then randomly throw out half of the selected objects in order to cut down on calculation time. The mean number of objects in the final photometric sample over the 24 light cones is $44,053$.

\begin{figure}[t]
\centering
\includegraphics[totalheight=\height]{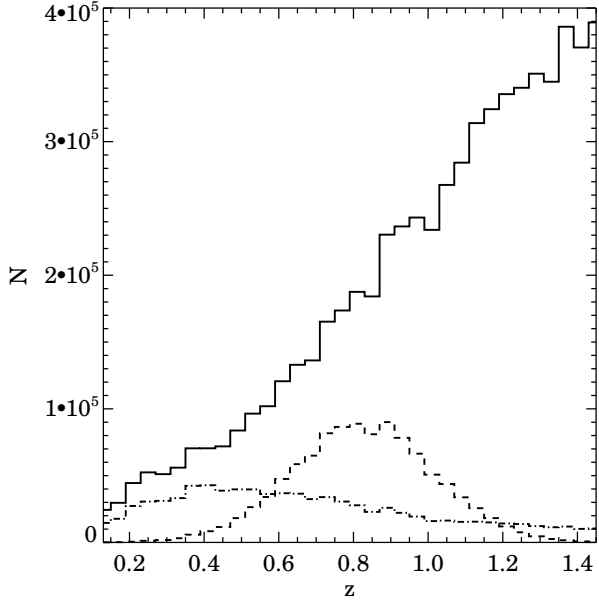}
\caption{The total number of galaxies in each sample as a function of redshift, summed over the 24 fields, binned with $\Delta z = 0.04$. The solid line is the overall redshift distribution for all galaxies in the mock catalogs, the dashed line is the distribution for our photometric sample (selected from the overall sample via a Gaussian in $z$, emulating objects placed in a single photometric redshift bin), while the dot-dashed line is the redshift distribution for our spectroscopic sample, selected to have magnitude $R<24.1$.}
\label{fig:ndist}
\end{figure}

The mock catalog includes both the cosmological redshift as well as the observed redshift for each object. The observed redshift shows the effects of redshift-space distortions \citep{1998ASSL..231..185H}, and is the redshift value used for objects in the spectroscopic sample. When plotting the redshift distribution of the photometric sample we use the cosmological redshifts for each object (differences are small). Fig. \ref{fig:ndist} shows the number of galaxies as a function of redshift for each sample, as well as the entire catalog.  While there is complete information on the actual redshift distributions for both samples in the catalog, only the distribution of the spectroscopic sample is assumed to be known in our calculations. We assume no information is known about the redshift distribution of the photometric sample, and attempt to recover it using only correlation measurements.

\section{Method}
\label{sec:method}
After constructing the two samples of objects from each mock catalog, we can use standard correlation measurements and exploit the clustering of galaxies to recover the redshift distribution of the photometric sample. From here on, the spectroscopic sample, with known observed redshifts, will be labeled '$s$', and the photometric sample, with redshifts assumed unknown, will be labelled '$p$'.

The most fundamental correlation measurements we use are the real space two-point correlation function and the angular two-point correlation function. The real space two-point correlation function $\xi(r)$ is a measure of the excess probability $dP$ (above that for a random distribution) of finding a galaxy in a volume $dV$, at a separation $r$ from another galaxy\citep{1980lssu.book.....P}: 
\begin{equation}
dP=n[1+\xi(r)]dV ,
\end{equation}
where $n$ is the mean number density of the sample. The angular two-point correlation function $w(\theta)$ is a measure of the excess probability $dP$ of finding a galaxy in a solid angle $d\Omega$, at a separation $\theta$ on the sky from another galaxy \citep{1980lssu.book.....P} :
\begin{equation}
dP=\Sigma[1+w(\theta)]d\Omega ,
\end{equation}
where $\Sigma$ is the mean number of galaxies per steradian (i.e., the surface density). From the spectroscopic sample we measure the real space two-point autocorrelation function, $\xi_{ss}(r,z)$, and from the photometric sample we measure the angular two-point autocorrelation function, $w_{pp}(\theta)$. These measurements give information about the intrinsic clustering of the samples. We also measure the angular cross\hyp{}correlation function between the spectroscopic and photometric sample, $w_{sp}(\theta,z)$, as a function of redshift. This is a measure of the excess probability of finding a photometric object at an angular separation $\theta$ from a spectroscopic object, completely analogous to $w_{pp}$.

Modeling $\xi(r)$ as a power law, $\xi(r)=(r/r_0)^{-\gamma}$, which is an accurate assumption from $\sim 0.5$ to $\sim 20 h^{-1}$ comoving Mpc for both observed samples and those in the mock catalogs, we can determine a relation between the angular cross\hyp{}correlation function $w_{sp}(\theta,z)$ and the redshift distribution. Following the derivation in \cite{2008ApJ...684...88N} (cf. eq. 4),
\begin{equation}
\label{eq:wsp}
w_{sp}(\theta,z) = \frac{\phi_p(z)H(\gamma_{sp})r^{\gamma_{sp}}_{0,sp}\theta^{1-\gamma_{sp}}D(z)^{1-\gamma_{sp}}}{dl/dz} ,
\end{equation}
where $H(\gamma)=\Gamma(1/2)\Gamma((\gamma-1)/2)/\Gamma(\gamma/2)$ (where $\Gamma(x)$ is the standard Gamma function), $\phi_p(z)$ is the probability distribution function of the redshift of an object in the photometric sample, $D(z)$ is the angular size distance, and $l(z)$ is the comoving distance to redshift $z$. Hence, to recover $\phi_p(z)$ from $w_{sp}$, we also must know the basic cosmology (to determine $D(z)$ and $dl/dz$), as well as the cross\hyp{}correlation parameters, $r_{0,sp}$ and $\gamma_{sp}$. It has been shown that uncertainties in cosmological parameters have minimal effect on the recovery of $\phi_p(z)$\citep{2008ApJ...684...88N}. To determine the cross\hyp{}correlation parameters, we use the assumption of linear biasing, under which the cross\hyp{}correlation is given by the geometric mean of the autocorrelations of the two samples, $\xi_{sp}(r)=(\xi_{ss}\xi_{pp})^{1/2}$. Thus we need to measure the autocorrelation functions for each sample and determine their parameters, $r_0$ and $\gamma$.

\subsection{Autocorrelation of the Spectroscopic Sample}
\label{sec:autospec}
We first need to determine how the real space autocorrelation function of the spectroscopic sample, $\xi_{ss}$, evolves with redshift. To do this we bin the spectroscopic objects in redshift and measure the two-point correlation function as a function of projected separation, $r_p$, and line-of-sight separation, $\pi$, for the objects in each bin. However, since it is affected by redshift-space distortions in the line of sight direction, it is difficult to measure the evolution of $\xi_{ss}(r)$ accurately directly from the observed $\xi(r_p,\pi)$. However, as we describe later, we can use $\xi(r_p,\pi)$ to derive the projected correlation function, $w_p(r_p)$, which is not significantly affected by redshift-space distortions. The evolution of the projected correlation function with redshift can be related to the evolution of $\xi(r)$. 

To begin we measure $\xi_{ss}$ in bins of $r_p$ and $\pi$, using the Landy \& Szalay estimator \citep{1993ApJ...412...64L}:
\begin{equation}
\label{eq:xi}
\xi=\frac{1}{RR}\!\left[DD\left(\!\frac{N_R}{N_D}\!\right)^{\!\!2}\!-2DR\left(\!\frac{N_R}{N_D}\!\right)+RR\right] ,
\end{equation}
where DD, DR, and RR are the number of object pairs in each bin of $r_p$ and $\pi$ -- i.e., the number of cases where an object of type B is located a separation of $r_p$ and $\pi$ away from an object of type A --  considering pairs between objects in the data catalog and other objects in the data catalog, between the data catalog and a random catalog, or within the random catalog, respectively; we will describe these catalogs in more detail shortly. Here $N_D$ and $N_R$ are the total numbers of objects in the data and random catalogs. For each object pair, we calculated the projected separation, $r_p$, and the line-of-sight separation, $\pi$, using the equations:
\begin{align}
\label{eq:rp}
r_p&=D(z_{mean})\Delta \theta  \\
\label{eq:pi}
{\rm and~~~}\pi&= |z_1-z_2|\ \frac{dl}{dz} \bigg|_{z_{mean}} ,
\end{align}
where $z_1$ and $z_2$ are the redshifts of the two objects in a pair, $\Delta \theta$ is their angular separation on the sky, and $z_{mean}=(z_1+z_2)/2$. 

We calculate DD by measuring the transverse and line-of-sight distance between every pair of objects in the data sample and binning those distances to find the number of pairs as a function of $r_p$ and $\pi$. In this case the data sample is all of the objects in the chosen spectroscopic $z$-bin. In turn, RR is the pair count amongst objects in a ``random'' catalog, and DR is the cross pair count calculated using pairs between data objects and random catalog objects. We construct the random catalog to have the same shape on the sky as the data catalog, but its objects are randomly distributed with constant number of objects per solid angle (taking into account the spherical geometry). 

To measure the real space correlation function, the random catalog must also have the same redshift distribution as the data catalog. To produce this, we first determine a smooth function that fits the overall redshift distribution of the spectroscopic sample and construct the random catalog to match. We had difficulty finding a single function that fit the entire distribution of $R<24.1$ galaxies in the Millennium mock from $z=0.1$ to $z=1.5$, so we used different functional forms over different redshift ranges. The best fit resulted from using $\phi_s(z) \sim z^2\exp(-z/z_o)$ for $0<z<1.03$ and $\phi_s(z) \sim A (1+z)^{\beta}$ for $z>1.03$. We bin the objects in each field into bins of $\Delta z=0.04$. Combining the distributions of all 24 fields and fitting via least-squares gave values of $z_o=0.232\pm0.003$ and $\beta=-2.74\pm0.18$. We then used these values, choosing a value of $A$ to force continuity at $z=1.03$, to define the redshift distribution used to generate the random catalogs. The random catalog for each field contained $\sim10$ times the number of objects as its corresponding data catalog.

After constructing the random catalogs, we calculate the pair counts in each redshift bin. For each field, both the data and random catalogs are divided into subsamples ("z-bins") according to their redshift, and DD, DR, and RR are calculated for each bin of $r_p$ and $\pi$ using only objects within a given z-bin. In the $r_p$ direction we binned the separations in $\log(r_p)$ over the range $-3 <\log(r_p)< 2.5$ with $\Delta\log(r_p)=0.1$, where $r_p$ is in $h^{-1}$Mpc. In the $\pi$ direction we binned the separations over the range $0 <\pi< 30\ h^{-1}$Mpc, with $\Delta\pi=1.0\ h^{-1}$Mpc. We calculated the pair counts in 10 z-bins covering the range $0.11<z<1.4$, where the size and location of each z-bin was selected so that there were approximately the same number of objects in each one.  

When interpreting correlation measurements for the spectroscopic sample, we must take into account the effects of redshift-space distortions \citep{1998ASSL..231..185H}. Since these only affect distance measurements along the line of sight, we integrate $\xi(r_p,\pi)$ in the $\pi$ direction, which gives the projected correlation function, $w_p(r_p)$. Modeling $\xi(r_p,\pi)$ as a power law and solving for $w_p(r_p)$ analytically gives 
\begin{align}
\label{eq:wp}
w_p(r_p)&=2\int^\infty_0\xi[(r^2_p+\pi^2)^{1/2}]d\pi  \\
\label{eq:wpa}
        &=r_p\left(\frac{r_0}{r_p}\right)^{\!\!\gamma} H(\gamma) ,
\end{align}
where $H(\gamma)$ is defined following equation \ref{eq:wsp}. We thus can recover $\gamma_{ss}(z)$ and $r_{0,ss}(z)$ by fitting a power-law model to $w_p(r_p)$ in each z-bin, allowing us to measure how the correlation function evolves with redshift. Because for our field geometry, signal-to-noise is poor at large scales, we fit for $w_p(r_p)$ up to $r_p=10\ h^{-1}$Mpc. The lower limit of $r_p$ used for the fit varied with redshift. We found in the highest redshift bins the behavior of $w_p(r_p)$ diverged from a power law, likely due to the semi-analytic model not populating group-mass halos with enough blue galaxies compared to DEEP2 data \citep{2008ApJ...672..153C}. Hence, for $z<0.8$ we fit over the range $0.1<r_p<10\ h^{-1}$Mpc, while for $z>0.8$ we fit over $1.0<r_p<10\ h^{-1}$Mpc.

We cannot measure $\xi(r_p,\pi)$ to infinite line-of-sight separations, so to calculate $w_p(r_p)$ we must integrate $\xi(r_p,\pi)$ out to $\pi_{max}=30\ h^{-1}$Mpc and then apply a correction for the fraction of the integral missed. In fact, in measuring $w_p(r_p)$, instead of evaluating $\xi(r_p,\pi)$ and then integrating, we simply summed the paircounts in the $\pi$ direction so DD, DR, and RR are functions of $r_p$ only; this method yielded more robust results. From equation \ref{eq:wp} (integrating to $\pi_{max}$ instead of infinity) we find
\begin{equation}
\label{eq:wp2}
w_p(r_p)= 2\left(\!\frac{1}{RR}\!\left[DD\!\left(\!\frac{N_R}{N_D}\!\right)^{\!\!2}\!-2DR\left(\!\frac{N_R}{N_D}\!\right)+RR\right]\right)\pi_{max},
\end{equation}
where DD, DR, and RR are the paircounts summed over the $\pi$ direction. For the correction, we first calculate $w_p(r_p)$ by summing the pair counts out to $\pi_{max}$, and then fit for $r_0$ and $\gamma$ using the analytic solution given in equation \ref{eq:wpa}. Using those parameters, we calculate $\int^{\pi_{max}}_0\xi(r_p,\pi)d\pi/\int^{\infty}_0\xi(r_p,\pi)d\pi$. We divide the observed $w_p(r_p)$ by this quantity and refit for $r_0$ and $\gamma$. This process is repeated until convergence is reached.

\subsection{Autocorrelation of the Photometric Sample}
\label{sec:autophot}
Since we assume the photometric sample contains no redshift information (or, more realistically, that any available redshift information was already exploited by placing objects into a redshift bin), we determine its autocorrelation parameters by measuring the angular autocorrelation function, $w_{pp}(\theta)$, and relating it to $r_{0,pp}$ using Limber's equation \citep{1980lssu.book.....P}:
\begin{equation}
\label{eq:wpp}
w_{pp}(\theta)=H(\gamma_{pp})\theta^{1-\gamma_{pp}}\!\!\int^\infty_0\!\!\phi^2_p(z)r^{\gamma_{pp}}_{0,pp}\frac{D(z)^{1-\gamma_{pp}}}{dl/dz}dz,
\end{equation}
where $\gamma_{pp}$ may be measured directly from the shape of $w_{pp}(\theta)$. We again measure the angular autocorrelation of the photometric sample using a Landy \& Szalay estimator:
\begin{equation}
w_{pp}(\theta)=\frac{1}{RR}\!\left[DD\left(\!\frac{N_R}{N_D}\!\right)^{\!\!2}\!-2DR\left(\!\frac{N_R}{N_D}\!\right)+RR\right] ,
\end{equation}
where DD, DR, and RR are the paircounts as a function of separation, $\theta$,  and $N_D$ and $N_R$ are the number of objects in the data and random catalogs for the field. For angular correlation measurements the random catalog consists of objects randomly distributed on the sky in the same shape as the data catalog. Again, the random catalog is $\sim10$ times larger than the data catalog. For each sample, we calculated the $\theta$ separation of every pair and binned them in $\log(\theta)$ over the range $-3 <\log(\theta)< 0.4$ with $\Delta\log(\theta)=0.1$, where $\theta$ is measured in degrees. 

The angular correlation function can be related to the spatial correlation function: $w_{pp}(\theta)=A_{pp}\theta^{1-\gamma_{pp}}$, where $A_{pp} \sim r^{\gamma_{pp}}_{0,pp}$ \citep{1980lssu.book.....P}. However, since the observed mean galaxy density in a field is not necessarily representative of the global mean density, our measurements of $w_{pp}(\theta)$ need to be corrected by an additive factor known as the integral constraint. To estimate this, we fit $w_{pp}(\theta)$ using a power law minus a constant, e.g. $w_{pp}(\theta)=A_{pp}\theta^{1-\gamma_{pp}}-C_{pp}$, where $C_{pp}$ is the integral constraint. For measuring the parameters we fit over the range $0.001^{\circ} <\theta< 0.1^{\circ}$. We found that fitting over this smaller range reduced the error in the amplitude measurements, although the error in the integral constraint (which is essentially a nuisance parameter) increases. For autocorrelation measurements this has little impact. We use the measured $\gamma_{pp}$, along with the parameters of the spectroscopic sample ($\gamma_{ss}(z)$ and $r_{0,ss}(z)$) and an initial guess of $r_{0,pp}$ to determine an initial guess of $r^{\gamma_{sp}}_{0,sp}$, employing the linear biasing assumption that  $r^{\gamma_{sp}}_{0,sp}=(r^{\gamma_{ss}}_{0,ss}r^{\gamma_{pp}}_{0,pp})^{1/2}$. 

We expect the correlation length of the photometric sample, $r_{0,pp}$, to be a function of redshift, as both the underlying dark matter correlation function and the large-scale structure bias of the sample will evolve with $z$, both in the real universe and in our mock catalogs. To account for this, we assume the redshift dependence of the scale length, $r_0$, will be similar for both the photometric and spectroscopic samples (we considered several alternatives, but this yielded the best results); for our calculations we set $r_{0,pp}(z) \propto r_{0,ss}(z)$, with an initial guess of $r_{0,pp}(z)=r_{0,ss}(z)$. We then refine our initial guess for $r^{\gamma_{sp}}_{0,sp}$ by measuring the angular cross\hyp{}correlation function in each redshift bin.

\subsection{Cross\hyp{}correlation and $\phi_p(z)$}
\label{sec:crossphi}
To find $w_{sp}(\theta,z)$, we measure the cross\hyp{}correlation between objects in spectroscopic z-bins with all objects in the photometric sample. We bin the spectroscopic sample over the range $0.19<z<1.39$ with a bin size of $\Delta z=0.04$ and measure $w_{sp}(\theta)$ for each bin using the estimator
\begin{align}
w_{sp}(\theta)=&\frac{1}{R_sR_p}\!\left[D_sD_p\!\left(\!\frac{N_{R_s}N_{R_p}}{N_{D_s}N_{D_p}}\!\right)\!-D_sR_p\!\left(\!\frac{N_{R_s}}{N_{D_s}}\!\right) \right. \nonumber \\
               &\left. -R_sD_p\!\left(\!\frac{N_{R_p}}{N_{D_p}}\!\right)+R_sR_p\right] ,
\end{align}
where $D_sD_p$, $D_sR_p$, $R_sD_p$, and $R_sR_p$ are the cross pair counts between samples as a function of $\theta$ separation, and $N$ is the number of objects in each sample. The cross pair counts are calculated by measuring the observed number of objects from one sample around each object in another sample. For example, $D_sD_p$ is the number of objects in the photometric sample around each spectroscopic object as a function of separation. For this measurement, each sample (the objects in the spec-z bin and the photometric sample) has their own random catalog that is $\sim10$ times bigger than their corresponding data catalog. These are once again constructed by randomly distributing objects on the sky in the same shape as the data catalog. 

For each z-bin we measured $w_{sp}(\theta)$ in logarithmic bins of 0.1 in $\log(\theta)$ over the range $-3 <\log(\theta)< 0.4$, with $\theta$ measured in degrees. As with the autocorrelation function, we fit $w_{sp}(\theta)=A_{sp}\theta^{1-\gamma_{sp}}-C_{sp}$; the integral constraint is nonnegligible in these measurements. Again we fit over the range $0.001^{\circ} <\theta< 0.1^{\circ}$ to reduce the error in the amplitude measurements. In some z-bins, particularly where the amplitude, $A_{sp}$, is small, we found a significant degeneracy between $A_{sp}$ and $\gamma_{sp}$ when fitting. One can understand this as there being a pivot scale at which clustering is best constrained; one can simultaneously vary $A_{sp}$ and $\gamma_{sp}$ and still match $w_{sp}$ at that scale. To remove this degeneracy, we fixed $\gamma_{sp}$ in each bin, and only fit for the amplitude and integral constraint. Since the clustering of the samples with each other is expected to be intermediate to the intrinsic clustering of each sample, we estimated $\gamma_{sp}$ with the arithmetic mean of $\gamma_{pp}$ and $\gamma_{ss}$. Using $A_{sp}$ and $\gamma_{sp}$, as well as the initial guess for $r^{\gamma_{sp}}_{0,sp}$, we determine an initial guess of the redshift distribution $\phi_p(z)$. Rewriting equation \ref{eq:wsp} gives
\begin{equation}
\label{eq:phi}
\phi_p(z) = \frac{dl/dz}{D(z)^{1-\gamma_{sp}}H(\gamma_{sp})r^{\gamma_{sp}}_{0,sp}}A_{sp}(z) .
\end{equation}
We then use the resulting  $\phi_p(z)$, along with $A_{pp}$ and $\gamma_{pp}$, to redetermine $r_{0,pp}$ using Equation \ref{eq:wpp}, which we use to redetermine $r^{\gamma_{sp}}_{0,sp}$ and thus $\phi_p(z)$. This process is repeated until convergence is reached.

\begin{figure*}[t]
\centering
\includegraphics[totalheight=\heightb]{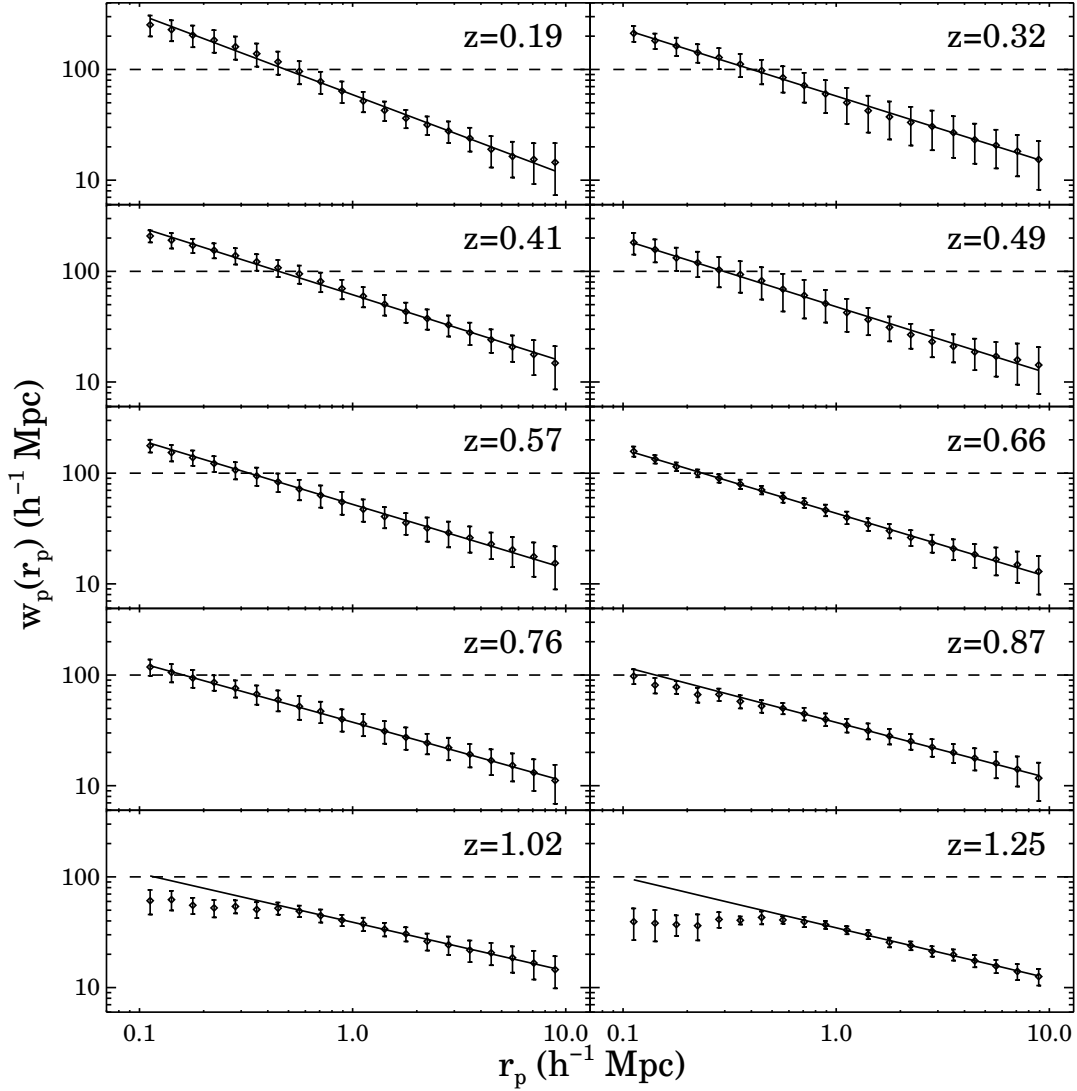}
\caption{The median value of $10^4$ measurements of the projected two-point correlation function of the spectroscopic sample, $w_p(r_p)$, in each redshift bin. Each measurement is made by averaging the paircounts of four fields selected at random from the 24 total fields. Error bars show the standard deviation of the measurements; i.e., they indicate the expected errors from a spectroscopic survey of four 1 square degree fields. The standard error in the plotted points is smaller than these error bars by a factor of $\sqrt{6}\ (2.45)$. At high redshift $w_p(r_p)$ deviates from a power law, whereas observed samples do not, due to the semi-analytic model not containing enough blue galaxies in group-mass halos. The solid line depicts a power-law model for $w_p(r_p)$, using the median values of the fit parameters $r_{0,ss}$ and $\gamma_{ss}$ across the $10^4$ measurements. The dashed line is the same in all panels; it is included to help make changes in the slope (i.e., $\gamma_{ss}$) and the amplitude (i.e., $r_{0,ss}$) with redshift clearer. We can see that changes in the amplitude with redshift are much more significant than changes in the slope.}
\label{fig:wprp}
\end{figure*}

\begin{figure}[t]
\centering
\includegraphics[totalheight=\height]{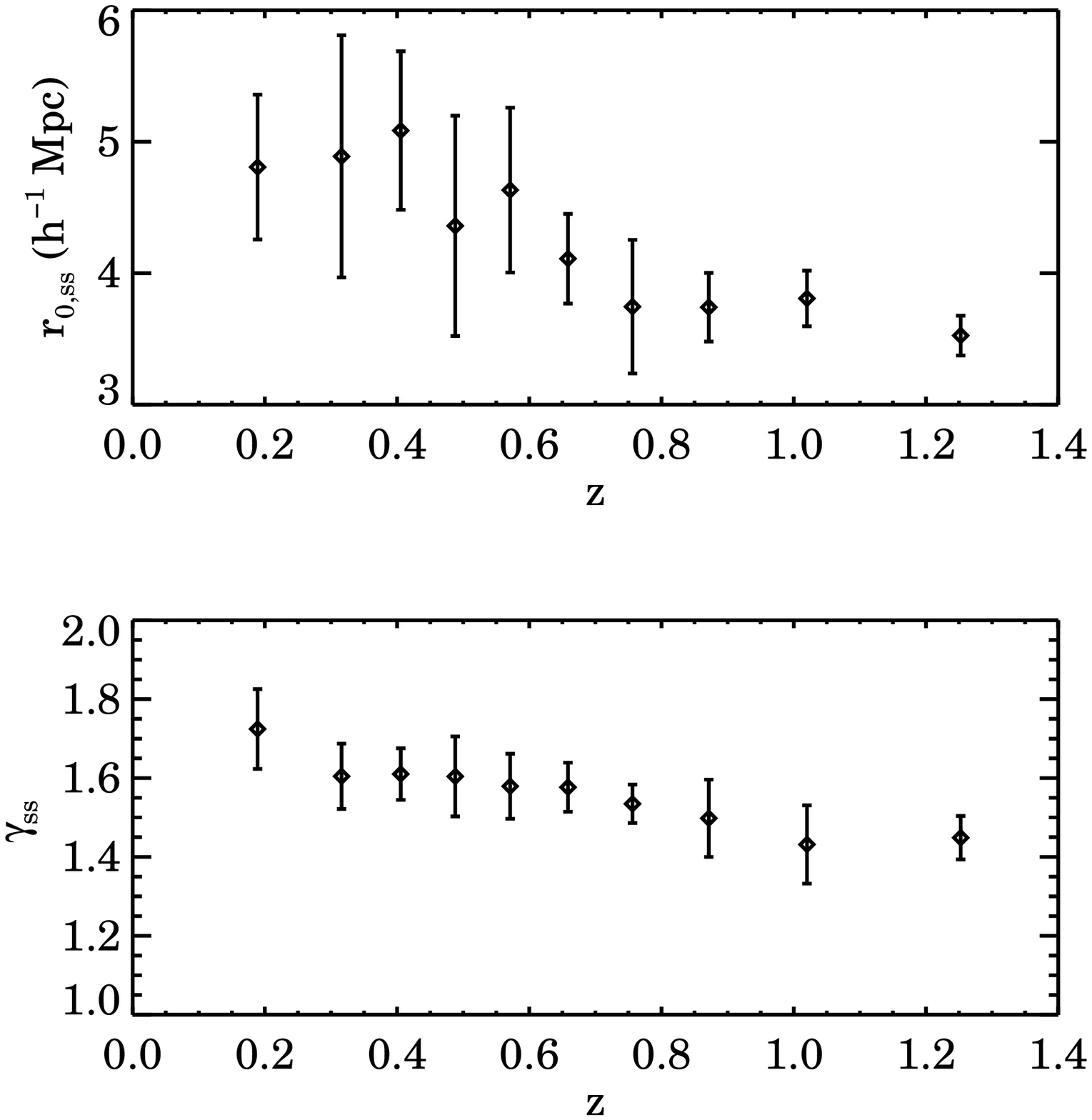}
\caption{The correlation function parameters resulting from power-law fits to $w_p(r_p)$, $r_{0,ss}$ and $\gamma_{ss}$, as a function of redshift. The points are the median values of $10^4$ measurements, and hence correspond to the parameters used to generate the lines in Fig. \ref{fig:wprp}; the error bars are the standard deviation of each parameter amongst the measurements. The standard error in the plotted points is smaller than these error bars by a factor of $\sqrt{6}\ (2.45)$. Each measurement is made by averaging the paircounts of four fields selected at random from the 24 total fields. While both parameters decrease with redshift, we see that changes in $r_{0,ss}$ are substantially greater than changes in $\gamma_{ss}$.}
\label{fig:rogm}
\end{figure}

\section{Results}
\label{sec:results}
For the remainder of the paper, we will frequently refer to making a ``measurement'' of the correlation functions and $\phi_p(z)$. Each measurement is done by selecting four fields at random out of the 24 mock catalogs, summing their pair counts, and calculating all necessary quantities; no information on 'universal' mean values of any measured quantity is used, but rather only that available from the chosen four fields. We select four fields in order to emulate redshift surveys like DEEP2 and VVDS, in which data is typically obtained from of order four separate fields; hence a 'measurement' in our parlance is roughly equivalent to utilizing the information coming from a single survey. To obtain the following results, we made $10^4$ measurements; we used the median values to evaluate statistical biases in a given quantity and the standard deviation to evaluate random uncertainties. In each plot following the points are the median values and the error bars are the standard deviations, which gives the error on a single measurement.  Because (given the large number of measurements) these medians should closely match the mean of the 24 fields, the standard error in a plotted point should be smaller than the plotted error bars by a factor of $\sqrt{6}$.

\begin{figure}[t]
\centering
\includegraphics[totalheight=\height]{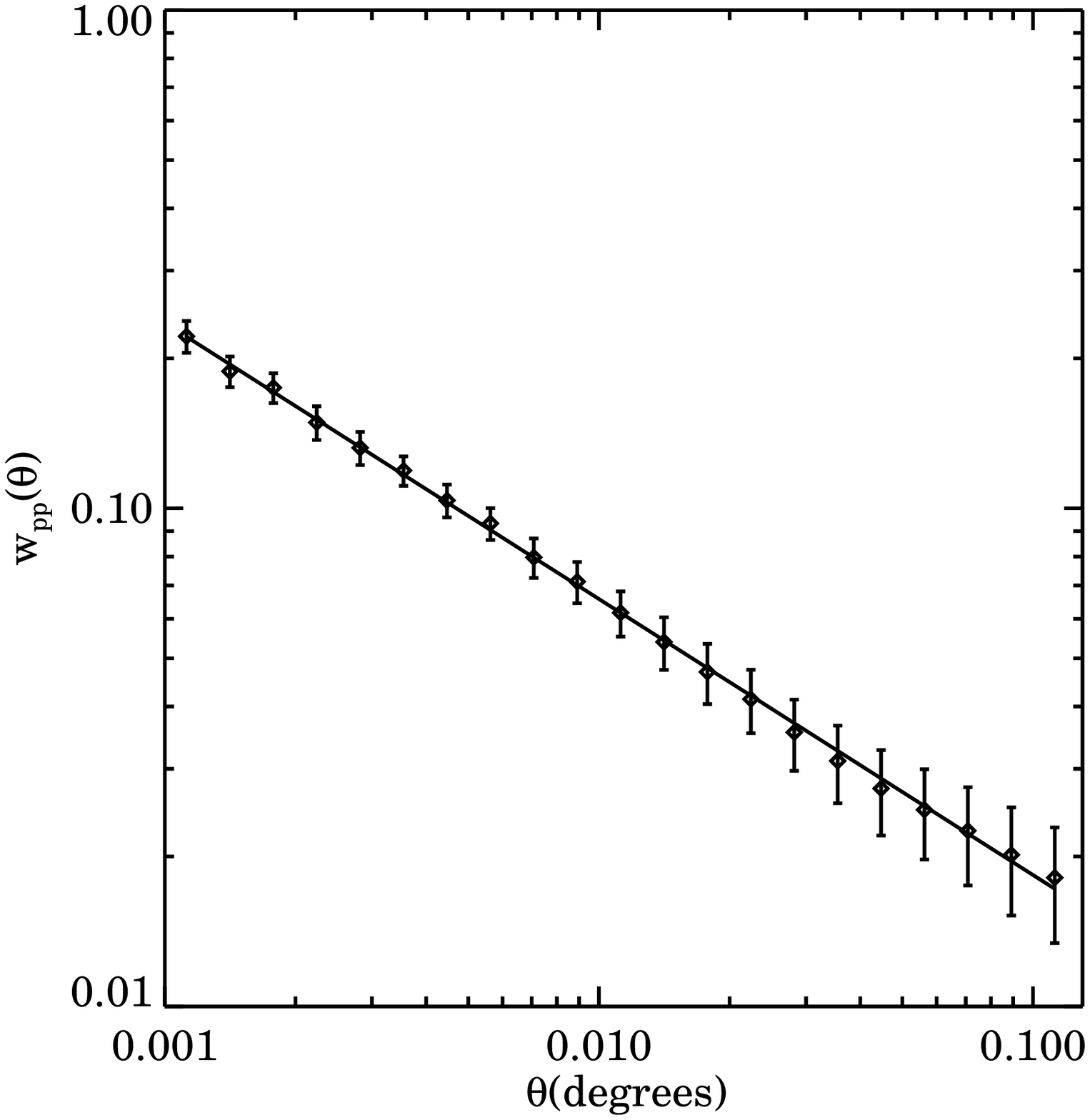}
\caption{The median value of $10^4$ measurements of the two-point correlation function of the photometric sample, $w_{pp}(\theta)$, corrected for the integral constraint. Each measurement is made by averaging the paircounts of four fields selected at random from the 24 mock catalogs. Error bars show the standard deviation of the measurements. The standard error in the plotted points is smaller than these error bars by a factor of $\sqrt{6}\ (2.45)$. The solid line is the fit to $w_{pp}(\theta)$ using the median values of the fit parameters $A_{pp}$ and $\gamma_{pp}$; a power-law model provides an excellent fit to the data.}
\label{fig:wpp}
\end{figure}

\begin{figure*}[t]
\centering
\includegraphics[totalheight=\heightb]{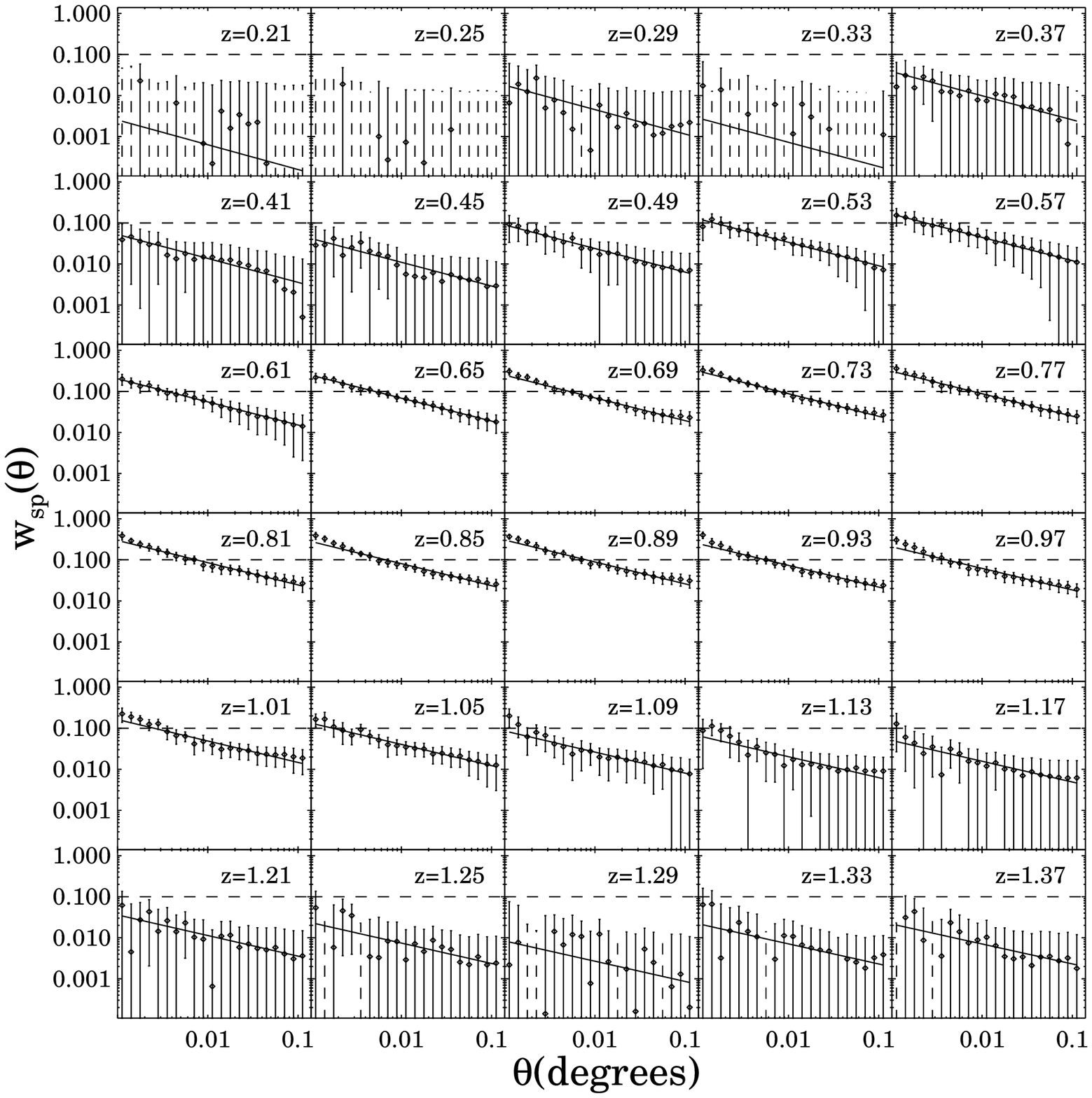}
\caption{The median value of $10^4$ measurements of the cross\hyp{}correlation between the photometric and spectroscopic samples, $w_{sp}(\theta)$, in each redshift bin, corrected for the integral constraint. Each measurement is made by averaging the paircounts of four fields selected at random from the 24 total fields. Error bars show the standard deviation of the measurements. The standard error in the plotted points is smaller than these error bars by a factor of $\sqrt{6}\ (2.45)$. The solid line is the fit to $w_{sp}(\theta)$ using the median values of the fit parameters $A_{sp}$ and $\gamma_{sp}$. The dashed line is to help make changes in the amplitude, $A_{sp}(z)$, with redshift clearer; in the fits shown the slope, $\gamma_{sp}(z)$, is forced to be constant with $z$. It is clear that the amplitude of the correlation is much greater in the central region of the redshift range where there are more photometric objects. The error on $w_{sp}(\theta)$ does not vary strongly with redshift, but rather the errors appear larger where there are few objects as a consequence of plotting on a logarithmic scale: i.e., the amplitude of the correlation is smaller in those regions, which leads to a much larger fractional error in $w_{sp}(\theta)$, and hence much larger error in log($w_{sp}$), where $w_{sp}$ is small, even though the error in $w_{sp}$ itself remains unchanged.}
\label{fig:wsp}
\end{figure*}

It should be noted that we are ignoring the weak cross correlation that should result from gravitational lensing by large-scale structure \citep{2008ApJ...684...88N,2010MNRAS.401.1399B}. These correlations can be predicted directly from galaxy number counts \citep{2005ApJ...633..589S}; planned surveys such as LSST will extend fainter than their nominal depth over limited regions of sky \citep{2009arXiv0912.0201L}, so no extrapolation will be required. It should also be possible to use the initial estimate of $\phi_p(z)$ to predict the lensing induced cross\hyp{}correlation signal at a given redshift, and therefore iteratively remove its contribution.  Because these correlation effects are weak, straightforward to deal with, and not present in the mock catalogs available to us, we do not consider them further here.

To determine the evolution of the autocorrelation parameters of the spectroscopic sample we measured $w_p(r_p)$ in z-bins of varying widths. Fig. \ref{fig:wprp} shows the median and standard deviation of $w_p(r_p)$ for $10^4$ measurements in each spectroscopic z-bin, with the correction for finite $\pi_{max}$ applied as described above. We then fit each measurement of $w_p(r_p)$ for the autocorrelation parameters.  The solid lines in Fig. \ref{fig:wprp} show the results of equation \ref{eq:wpa} corresponding to the median $r_{0,ss}$ and $\gamma_{ss}$ for all measurements in a given z-bin, while Fig. \ref{fig:rogm} shows the accuracy with which we can measure the evolution of $r_{0,ss}$ and $\gamma_{ss}$ with redshift. Both parameters decreasing with redshift is consistent with measurements in real samples which show bluer galaxy samples have smaller $r_0$ and $\gamma$ \citep{2008ApJ...672..153C}; a constant observed magnitude limit will correspond to a selection at a bluer and bluer rest frame band as redshift goes up, increasingly favoring bluer objects for selection.

The autocorrelation parameters for the photometric sample are determined from the shape of $w_{pp}(\theta)$.  Fig. \ref{fig:wpp} shows the median and standard deviation of $10^4$ measurements of $w_{pp}(\theta)$, corrected for the integral constraint. A fit to each measurement gives estimates of autocorrelation parameters. Taking the median values and standard deviations gives $A_{pp}=5.48\times10^{-4}\pm2.73\times10^{-4}$ and $\gamma_{pp}=1.55\pm0.045$. The solid line in Fig. \ref{fig:wpp} corresponds to these median values. The scale length of the photometric sample, $r_{0,pp}(z)$, was assumed to be proportional to $r_{0,ss}(z)$; this yielded superior results to other simple assumptions. The proportionality constant may then be found using an initial guess of $r_{0,pp}=r_{0,ss}$ to calculate $\phi_p(z)$ using cross\hyp{}correlation techniques, leading to a refined estimate of $r_{0,pp}$ using Limber's equation (eqn. \ref{eq:wpp}).  That refined $r_{0,pp}$ is then used to make an improved measurement of $\phi_p(z)$, which is used to obtain a yet-improved measure of $r_{0,pp}$, etc.  After convergence was reached, we found that on average $r_{0,pp}/r_{0,ss}=1.068$.

\begin{figure}[t]
\centering
\includegraphics[totalheight=\height]{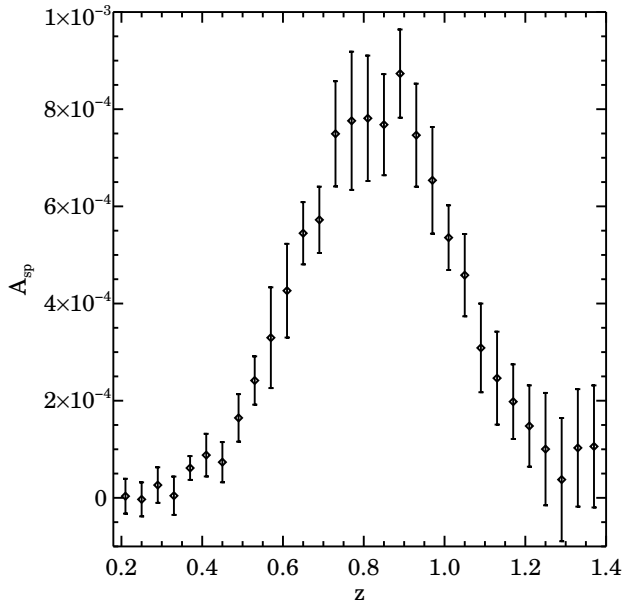}
\caption{The median value of $10^4$ measurements of $A_{sp}$, the amplitude of $w_{sp}$, in each redshift bin.  Each plotted point corresponds to the amplitude of one of the model lines shown in Fig. \ref{fig:wsp}.  Each measurement is made by averaging the paircounts of four fields selected at random from the 24 mock catalogs. Error bars show the standard deviation of the measurements. The standard error in the plotted points is smaller than these error bars by a factor of $\sqrt{6}\ (2.45)$. The amplitude is larger in the central region of the redshift range where there are more photometric objects, which is expected since the degree to which the two samples overlap in redshift contributes to the strength of the cross\hyp{}correlation function.}
\label{fig:asp}
\end{figure}

To determine the evolution of the cross\hyp{}correlation parameters, we measure the angular cross\hyp{}correlation, $w_{sp}(\theta, z)$, between objects in successive spectroscopic z-bins and the photometric sample. Fig. \ref{fig:wsp} shows the median and standard deviation of $w_{sp}(\theta)$ for $10^4$ measurements in each z-bin, corrected for the integral constraint. Fitting each measurement for the cross\hyp{}correlation parameters with fixed $\gamma_{sp}$ as described above and taking the median gives the amplitude, $A_{sp}(z)$, shown in Fig. \ref{fig:asp}. The solid lines in Fig. \ref{fig:wsp} correspond to the median of the best-fit parameters from each measurement. 

Combining the intrinsic clustering information from the autocorrelation parameters of each sample with the amplitude of the cross\hyp{}correlation, $A_{sp}(z)$, together with the basic cosmology, gives the recovered redshift distribution. We found that a linear fit of $r_{0,ss}$ and $\gamma_{ss}$ versus z resulted in a better recovery of $\phi_p(z)$ than using each bin's value directly, resulting in a $\sim32\%$ reduction in the $\chi^2$ of the final reconstruction as compared to the true redshift distribution. Fitting the correlation function over a limited $\theta$ range, as described in $\S$ \ref{sec:crossphi}, reduced the measured error in $\phi_p(z)$ for each z-bin by $\sim25\%$ on average, reducing the $\chi^2$ in comparing the reconstructed and true redshift distributions by $\sim30\%$. We also tried modeling $\gamma_{sp}$ as constant with z using the arithmetic mean of $\gamma_{ss}(z=0.77)$ and $\gamma_{pp}$. This resulted in a $\sim20\%$ increase in the $\chi^2$ of the final fit.

\begin{figure}[t]
\centering
\includegraphics[totalheight=\height]{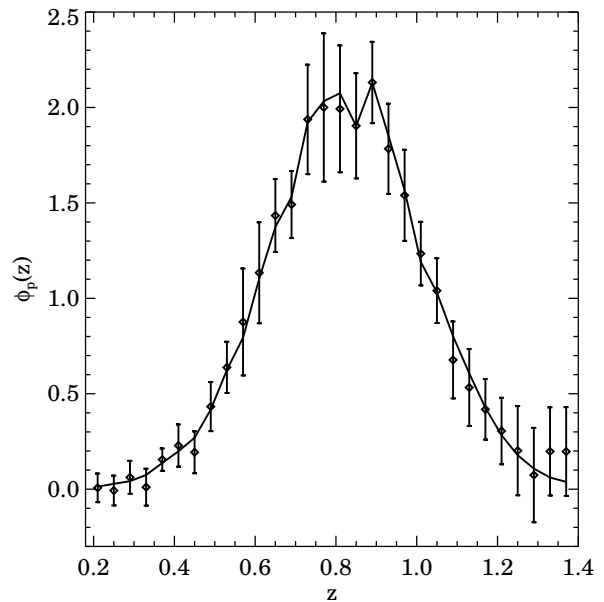}
\caption{Plot of the redshift distribution recovered using cross\hyp{}correlation techniques. The solid line is the actual distribution of the photometric sample (combining all 24 fields), while the points are the median reconstructed values from $10^4$ measurements. Error bars show the standard deviation of the recovered distribution when performing cross\hyp{}correlation reconstruction in 4 $0.5 \times 2$ deg fields, emulating the data available from existing deep redshift surveys. The standard error in the plotted points is smaller than these error bars by a factor of $\sqrt{6}\ (2.45)$. Each measurement is made by averaging the paircounts of four fields selected at random from the 24 mock catalogs. The recovered distribution follows the true distribution closely, even picking up the irregular dip due to sample variance (also known as cosmic variance) at the peak.}
\label{fig:phi}
\end{figure}

Fig. \ref{fig:phi} shows the median and standard deviation of $10^4$ measurements of $\phi_p(z)$ compared to the actual distribution. To determine the actual distribution, we found the mean true distribution of the four fields corresponding to each measurement and took the median across the $10^4$ measurements; this should accurately match the true mean of the redshift distributions over the 24 fields. Each measurement was normalized so that integrating $\phi_p(z)$ over the measured redshift range gives unity before the median was taken. It is important to note that the reconstruction techniques we have implemented thus far will recover the actual redshift distribution of objects in the photometric sample.  This will in general deviate from the true, universal redshift distribution of objects of that type due to sample/cosmic variance. We describe and test methods for recovering the underlying universal distribution in $\S\ref{sec:errorest}$.

\begin{figure*}[t]
\centering
\includegraphics[totalheight=\heightb]{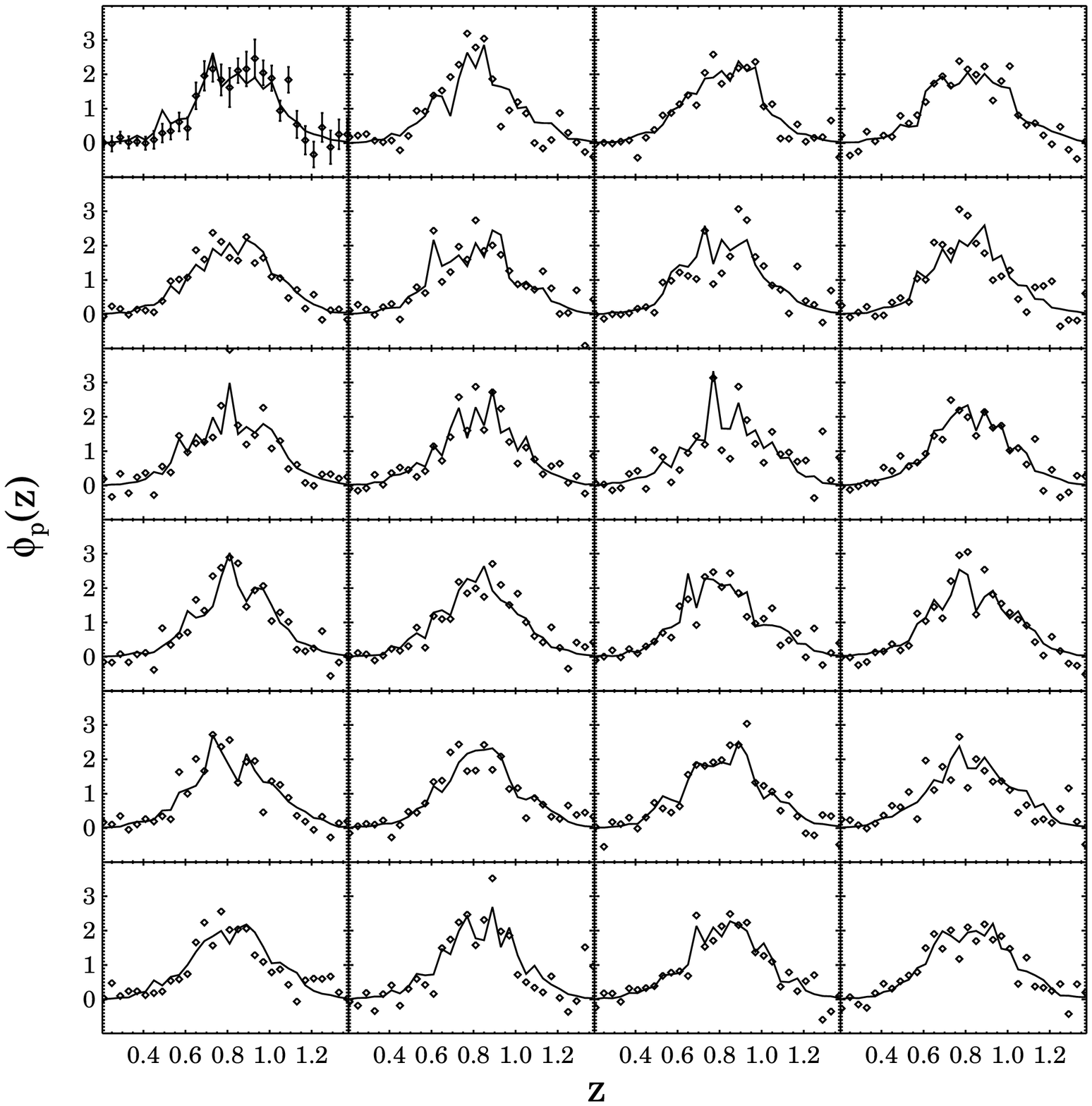}
\caption{Plot of the recovered redshift distribution for each of the 24 fields, using only pair counts from a single field in the reconstruction. The error bars in the first plot are the standard deviation of $\phi_{p,rec}(z)-\phi_{p,act}(z)$ amongst the 24 fields; they should be representative of the expected error for each panel. For each field, all errors used in fitting are based on standard deviations across the 24 fields. This mimics a common situation where we have only one field, but use errors determined from simulations to weight points for fitting.  The reconstruction generally captures the variation amongst fields due to sample/cosmic variance.}
\label{fig:phising}
\end{figure*}

We also looked at how well redshift distributions may be recovered in a single, 1 square degree field. For each field, the correlation functions were calculated using only the information from that field. To weight each bin when fitting for correlation-function parameters, the fit was calculated using errors given by the standard deviation of the correlation function in each $\theta$ bin over the 24 fields. This mimics the common situation where we have few fields with data and errors are determined from simulations. For a single field, a linear fit for the evolution of the spectroscopic-sample correlation function parameters was not a good model, so we used the calculated parameters in each z-bin. Fig. \ref{fig:phising} shows the recovered distribution, $\phi_p(z)$, in each of the 24 fields, compared to the true redshift distribution of the photometric sample in that field. 

\begin{figure}[t]
\centering
\includegraphics[totalheight=\height]{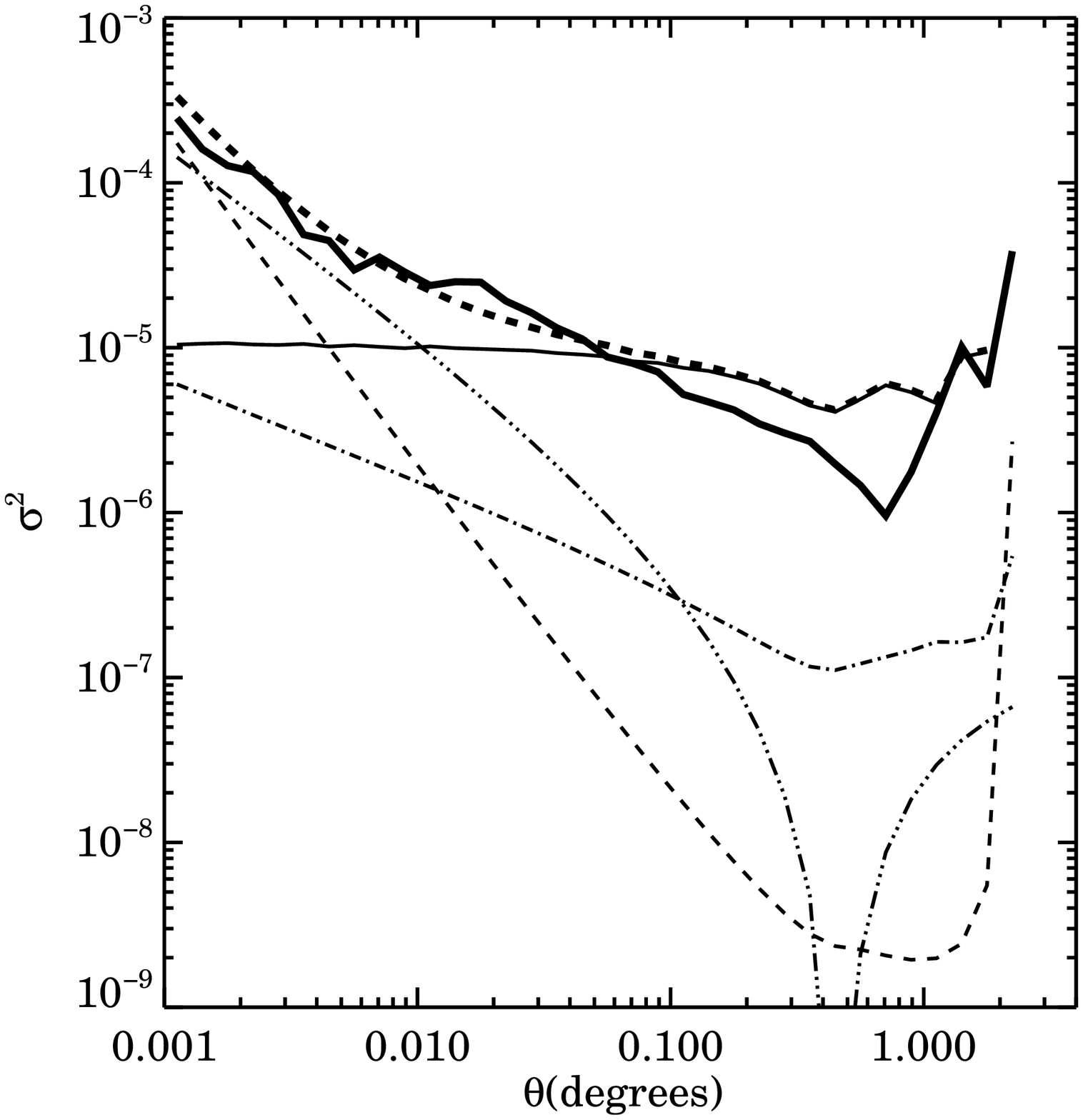}
\caption{The variance of $10^4$ measurements of the autocorrelation of the photometric sample, $w_{pp}(\theta)$ (thick solid line), compared to predicted error terms from Bernstein 1994. The thick dashed line shows the sum of all the variance terms; it corresponds well to the observed variance save at the largest scales, where the Bernstein 1994 model is overly conservative (a consequence of the assumption made in that work that the angular separations considered are significantly smaller than the size of the field). From equation 38 in \cite{1994ApJ...424..569B}, the thin solid black line is the term that scales as $w^2$, corresponding to the variance in the integral constraint, which dominates at large $\theta$. The thin three-dot-dash line is the term that scales at $w^3$, and the thin dot-dash line is the term that scales as $\mathrm{1/N}$. The thin dashed black line is the term that scales as $\mathrm{1/N^2}$ and is comparable to the Poisson error, which dominates in the weak clustering formalism used by Newman (2008). The 'observed' variance in $w_{pp}(\theta)$ is much larger than the weak clustering prediction; the same is true of $w_{sp}(\theta)$, although to a lesser degree.}
\label{fig:errors}
\end{figure}

\subsection{Correlation Measurement Errors}
\label{sec:correrrors}
In the course of our calculation of the redshift distribution, we found that the error in $\phi_p(z)$ for each redshift bin was larger than expected from the error model used in \cite{2008ApJ...684...88N}, which uses the standard, classical weak-clustering formalism. This formalism predicts that Poisson uncertainties should dominate when the clustering strength (e.g. the value of $w_{sp}$) is small compared to unity \citep{1980lssu.book.....P}. Upon further investigation we determined that the error in all correlation function measurements were larger than expected according to this model, which led to the excess error in $\phi_p(z)$. This additional error is associated with extra variance terms identified by \cite{1994ApJ...424..569B}, which contribute significantly even in the weak-clustering limit, contrary to the classical assumption. These extra terms are dominated by the variance in the integral constraint, which has a significant impact if spectroscopic samples cover only a few square degrees of sky.

Fig. \ref{fig:errors} compares the four terms of the predicted error from Bernstein's error model to our measured error for $w_{pp}(\theta)$. Bernstein's error model assumes the separation is much smaller than the field size, so we see for small $\theta$ the predicted variance does follow our measured variance closely, and then deviates as the separation becomes comparable to the field size. The integral constraint term dominates at large $\theta$ values. In order to calculate some of the variance terms of Bernstein's model we required values for $q_3$ and $q_4$, which are used to relate the three- and four-point correlation functions to the two-point correlation function assuming hierarchical clustering. For this we used the values measured by Bernstein in simulation catalogs, $q_3=0.32$ and $q_4=0.1$ \citep{1994ApJ...424..569B}. This gave a better fit to our results than the values observed in local galaxy samples \citep{1992ApJ...394...87M,1992ApJ...390..350S}.

From Fig. \ref{fig:errors} we see that the measured variance can be orders of magnitude larger than errors predicted using the weak-clustering assumption (though the difference is a smaller factor for $w_{sp}$, whose errors dominate in reconstructing $\phi_p(z)$). This excess variance will have a significant impact on the error budgets of planned dark energy experiments (see the next section for quantitative estimates); it is dominated by the variance in the integral constraint, whose effect increases with decreasing field size, so errors may be greatly reduced by surveying galaxies over a larger area ($>\sim 100$ square degrees instead of $\sim 4$).  For instance, the proposed BigBOSS survey\citep{2009arXiv0904.0468S} would provide a near-ideal sample for cross\hyp{}correlation measurements (using both galaxies and Lyman $\alpha$ absorption systems at redshifts up to $\sim 3$). We may also reduce this effect by using better correlation function estimators which reduce the effect of the integral constraint. One example of such a robust estimator relies on convolving the two-point correlation function with a localized filter \citep{2007MNRAS.376.1702P}. We are currently testing the $\phi_p(z)$ reconstruction using this estimator to determine its impact on the error in our recovered redshift distribution.

\begin{figure*}[t]
\centering
$\begin{array}{@{\hspace{-.2in}}c@{\hspace{-.2in}}c}
\includegraphics[totalheight=\height]{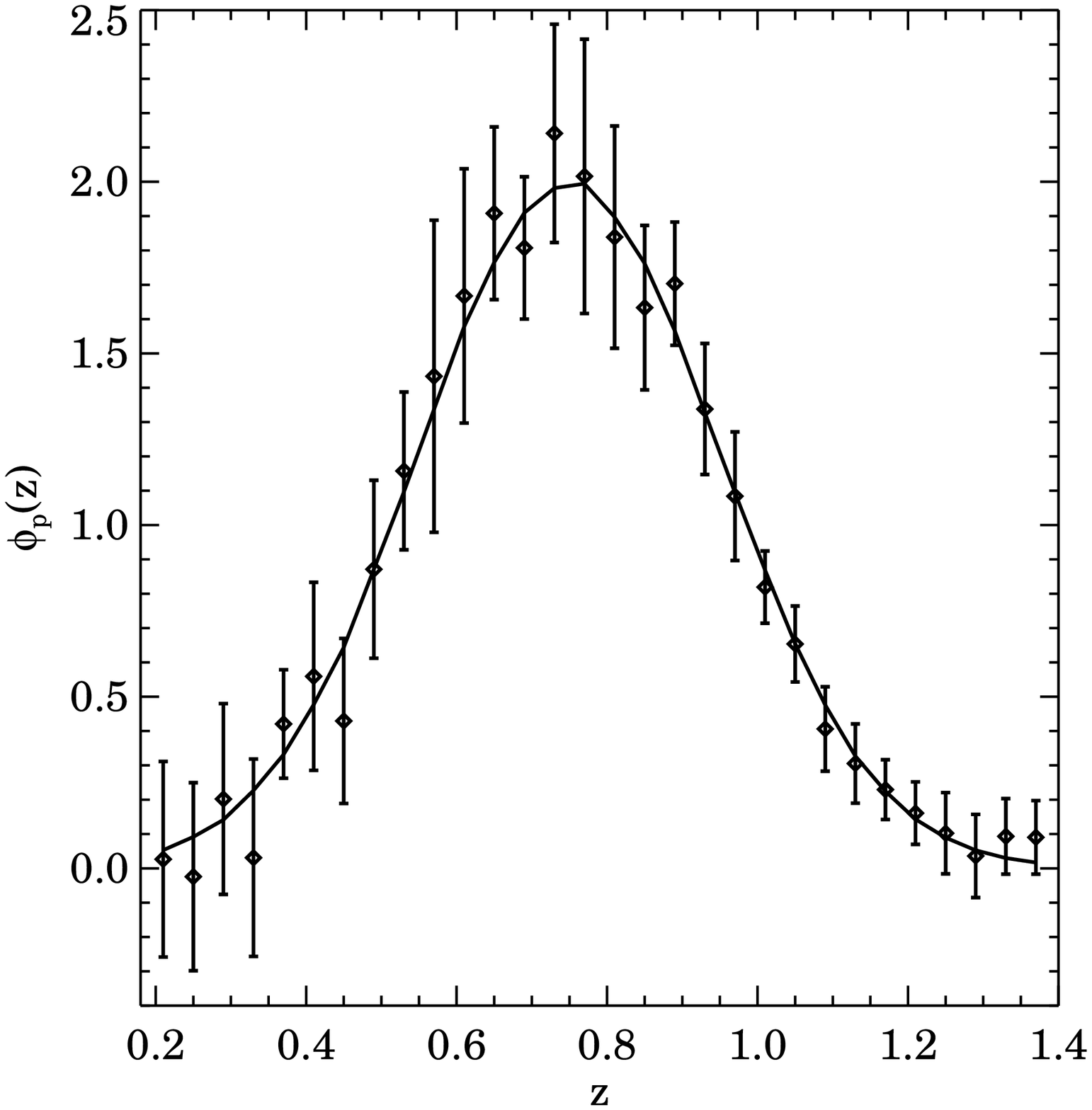}  &
\includegraphics[totalheight=\height]{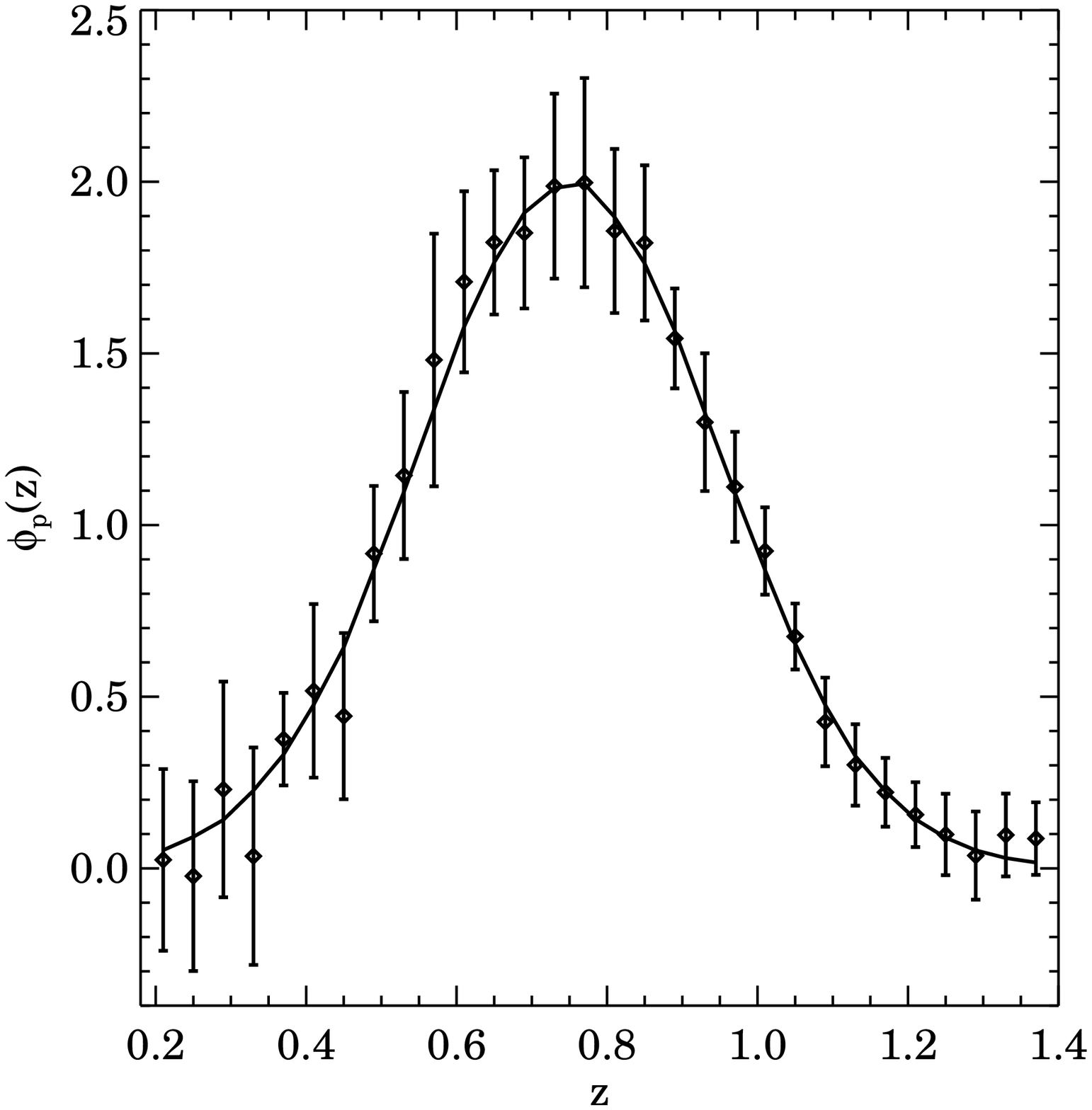} \\
\end{array}$
\caption{Plots of the recovered and mean true redshift distribution of the 24 fields, after the overall redshift distribution of all galaxies in the mock catalogs, $dN/dz$, is divided out, as described in $\S\ref{sec:errorest}$. On the left is the reconstruction before applying a correction for sample/cosmic variance based on fluctuations in the spectroscopic redshift distribution in the fields observed, and on the right is the reconstruction after that correction. There is a significant improvement in the reconstruction. The plot on the right corresponds to the reconstruction of the probability an object falls in the photometric redshift bin as a function of its true $z$ (or, equivalently, the reconstruction of the photometric redshift error distribution), rather than reconstructing the actual redshift distribution (affected by sample/cosmic variance) of galaxies in a particular set of fields, as was depicted in Fig. \ref{fig:phi}.  The solid line in each panel is the true normalized distribution of the photometric sample and the points are the median values of $10^4$ measurements. The true distribution matches the Gaussian selection function used for creating the photometric sample, by construction. Error bars show the standard deviation of the recovered distribution. The standard error in the plotted points is smaller than these error bars by a factor of $\sqrt{6}\ (2.45)$. Each measurement is made by averaging the paircounts of four fields selected at random from the 24 total mock catalogs available.  As shown here, if we know the amplitude of fluctuations from cosmic variance at a given redshift (using the variance in the distribution of spectroscopic galaxies), as well as the overall distribution of the parent sample (e.g. from combining redshift distributions from all photometric redshift bins), we can accurately reconstruct the true selection probability distribution.}
\label{fig:phicvar}
\end{figure*}

\begin{figure}[t]
\centering
\includegraphics[totalheight=\height]{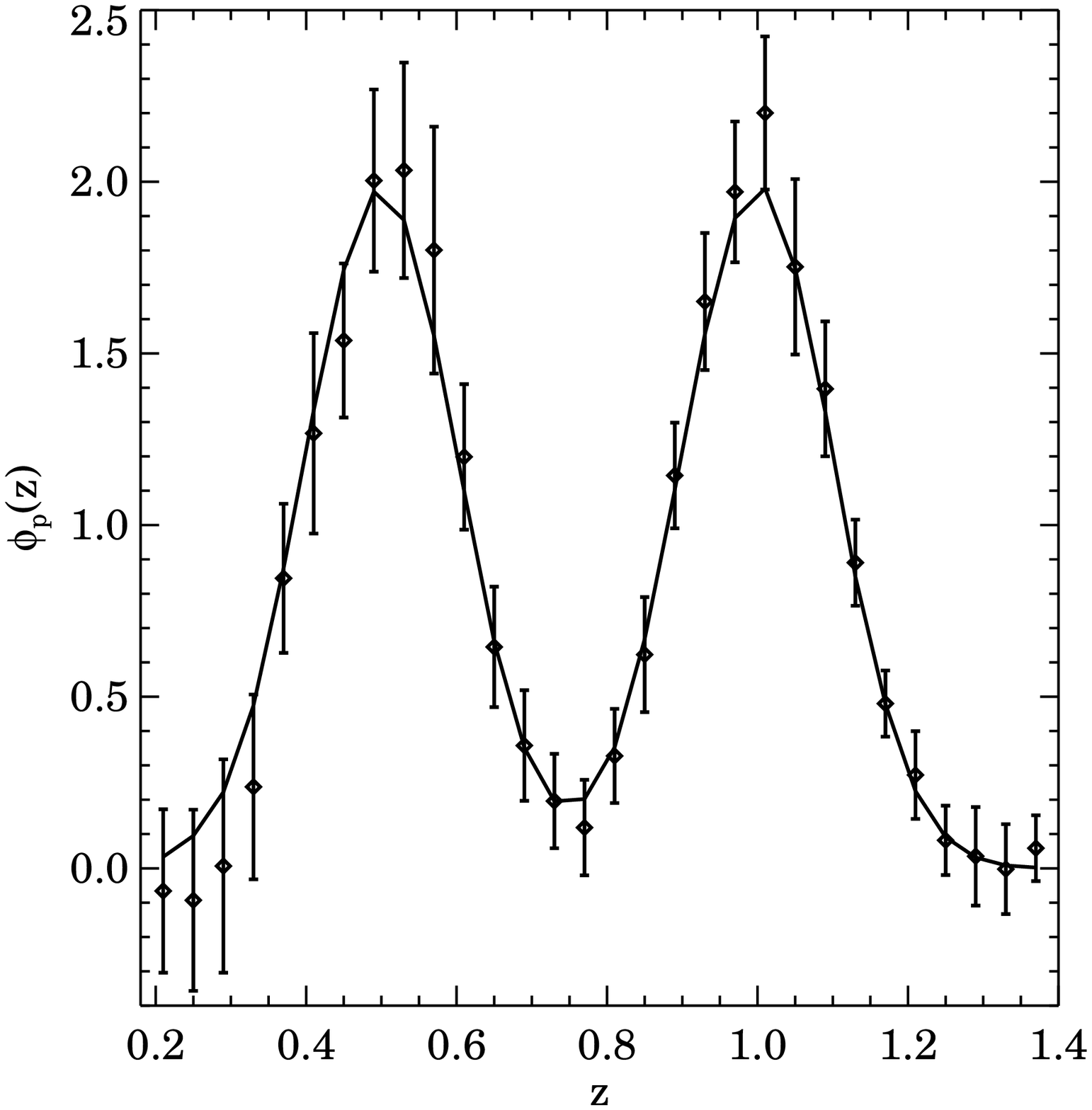}
\caption{Results of cross\hyp{}correlation reconstruction of a selection function consisting of two equal-amplitude Gaussian peaks centered at $z=0.5$ and $z=1.0$, each with $\sigma_z=0.1$. The solid line is the true distribution of the photometric sample (combining all 24 fields), while the points are the median reconstructed values from $10^4$ measurements. Error bars show the standard deviation. The standard error in the plotted points is smaller than these error bars by a factor of $\sqrt{6}\ (2.45)$. Each measurement is made by averaging the paircounts of four fields selected at random from the 24 mock catalogs. This plot is analogous to the right panel of Fig. \ref{fig:phicvar}; as in that case, we are reconstructing the selection function of the sample rather than its redshift distribution. The effects of bias evolution should be greater in this case, however, as the sample is less concentrated in redshift. The recovery remains accurate here, despite the larger bias evolution and very different $\phi_p(z)$.}
\label{fig:phi2peak}
\end{figure}

\subsection{Error Estimates}
\label{sec:errorest}
In this subsection, we investigate the impact of these excess correlation function measurement errors on our ability to recover the parameters (i.e. the mean and $\sigma$) of the true redshift distribution for the photometric sample, and compare the results to Monte Carlo tests done in \cite{2008ApJ...684...88N}. For each measurement we have a recovered distribution and an associated true distribution for that set of four fields. We will test the recovery both of the underlying, universal distribution used to construct the photometric sample (i.e. $\langle z \rangle=0.75$, $\sigma_z=0.20$) and of the actual redshift distribution of the objects selected in a given set of fields (which will differ due to sample/cosmic variance; cf. $\S\ref{sec:results}$). 

Before we can fit for Gaussian parameters, we must account for the fact that our photometric sample has a redshift distribution which differs from a true Gaussian because the total sample we drew from (with Gaussian probability as a function of $z$) was not uniformly distributed in redshift. One can think of the actual distribution of the photometric sample in a given bin as a product of three factors: the overall redshift distribution of all objects in the Universe (essentially, the rising curve in \ref{fig:ndist}); the fractional deviation from the Universal mean of the number of objects in a given field at a given redshift, i.e. sample/cosmic variance; and the Gaussian function used to select objects for the photometric redshift bin. 

The first two factors need to be removed from both the true and recovered distributions if we are to test the recovery of the third; this is implemented differently for each case. For the true distribution, we divide each measurement by the overall $dN/dz$ of all of the objects in the four fields used in that measurement. This removes the overall distribution shape as well as the fluctuations due to sample variance, and gives a true distribution that closely matches the Gaussian selection function applied to construct the sample. 

In principle we could do the same for the recovered distribution, but that would not be practical in real applications, as we can determine the overall shape of the redshift distribution of the overall photometric sample using photometric redshifts, but photo-z errors will prevent measuring fluctuations in the number of objects within bins of small $\Delta z$.  Hence,
we correct the recovered $\phi_p(z)$ using a low-order polynomial fit to the shape of the overall sample's $dN/dz$, but use the fluctuations (compared to a smooth fit) in the observed redshift distribution of the spectroscopic sample $dN_s/dz$, which will be known from the same observations used to perform cross\hyp{}correlation measurements, to correct for sample variance. This correction assumes that deviations from the mean in both samples behave similarly with redshift; we might expect their amplitude to scale with the large-scale-structure bias of a given sample, but we do not apply any correction for that here. In tests, we have found that a correction using fluctuations in $dN_s/dz$ was as effective in constraining parameters as one based on fluctuations in the $dN/dz$ of the overall sample our photometric subsample was selected from, and so we focus on the former, more realistic technique.

In more detail, we first divided the recovered distribution by a smooth fit (using a 5th-degree polynomial function) to the overall $dN/dz$ of the entire simulation averaged over all 24 fields. This eliminates gradients associated with the shape of the parent sample's overall redshift distribution without removing deviations due to sample variance. To correct for the latter, we need to quantify the fluctuations in the spectroscopic sample relative to a mean distribution. For this smooth, mean distribution, $\langle dN_s/dz\rangle$, we used the same fit to the redshift distribution of the spectroscopic sample averaged over all 24 fields which was employed to construct the random catalogs for autocorrelation measurements ($\S\ref{sec:autospec}$).  Using a fit to a given set of four fields would make little difference, as the deviation from the smooth fit at a given redshift bin due to sample variance are much larger than the deviations between the smooth fit to 4 or 24 fields. We then calculate the ratio $dN_s/dz/\langle dN_s/dz\rangle$, where $dN_s/dz$ is the redshift distribution of the spectroscopic sample averaged over the four fields used in that measurement, and correct for sample variance by dividing each measurement of $\phi_p(z)$ by this quantity.

After applying these corrections to each distribution, each measurement is normalized so that their integral is unity, and then fit for $\langle z \rangle$ and $\sigma_z$ using a normalized Gaussian fitting function. Fig. \ref{fig:phicvar} shows the median and standard deviation of $10^4$ measurements of the recovered $\phi_p(z)$ before and after correcting for sample variance. In both plots the fit to the overall $dN/dz$ is divided out. It is clear to the eye that the distribution corrected for sample variance is a better fit to the underlying selection function; more quantitatively, it reduces errors in determining the parameters of the Gaussian selection function by $\sim10\%$.

We assess the reconstruction of the photometric sample in two ways. First, we compare the reconstructed parameters, $\langle z \rangle$ and $\sigma_z$, of the Gaussian selection function to the true values, known by construction. Second, we compare the reconstructed parameters of the selection function to the parameters of a Gaussian fit to the actual normalized distribution of each set of four fields used. The latter method should be more robust to systematic errors in the 'true' $dN/dz$ we divide each measurement by.

For the first test, where $\langle z\rangle_{true} = 0.75$ and $\sigma_{z,true} = 0.20$, we find $\langle \langle z \rangle_{rec} -  \langle z \rangle_{true}\rangle = 7.796\times10^{-4}\pm7.415\times10^{-3}$ and $\langle \sigma_{z,rec} - \sigma_{z,true} \rangle = 8.140\times10^{-4}\pm8.545\times10^{-3}$, where as usual the values given are the median and standard deviation of all measurements, respectively. The second test, where $\langle z\rangle_{true}$ and $\sigma_{z,true}$ are determined by a Gaussian fit to the true distribution of each measurement, we find $\langle \langle z\rangle_{rec}-\langle z\rangle_{true}\rangle = 7.259\times10^{-4}\pm7.465\times10^{-3}$ and $\langle \sigma_{z,rec}-\sigma_{z,true}\rangle = 4.724\times10^{-4}\pm8.546\times10^{-3}$. In all cases, the bias is not statistically significant (the standard error against which each bias estimate must be compared is smaller than the quoted standard deviations by a factor of $\sqrt{6}$), but in any event the overall bias of both parameters is considerably smaller than the associated random errors, and will therefore have little effect when added in quadrature. These errors are still larger than the estimated requirements for future surveys (i.e. $\sigma \sim 2-4\times10^{-3}$, as described in \S\ref{sec:intro}). For cross\hyp{}correlation techniques to meet these requirements, this excess error will need to be reduced. We discuss a few options for this in $\S\ref{sec:correrrors}$.

A number of choices we have made on how to model and measure correlation function parameters (e.g. using a fit for the dependence of the spectroscopic sample's autocorrelation parameters on z vs. using the values for a given z-bin directly; assuming $r_{0,pp} \propto r_{0,ss}$ vs. a constant $r_{0,pp}$; or allowing $\gamma_{sp}(z)$ to decrease with redshift vs. forcing a constant $\gamma_{sp}$) can affect both the bias and error in these measurements. We have tested reconstruction with alternate methods to those described here and found that the random errors in $\langle z \rangle$ and $\sigma_z$ are much more robust to these changes than the bias. When varying the three correlation parameters as described previously, the standard deviation of the measurements never varied by more than $\sim10\%$, but the bias in some cases increased significantly. For measurements of $\langle z \rangle$, the alternative parameter models yielded biases of $0.006-0.009$, making them statistically significant compared to the random errors. For $\sigma_z$, the biases under the different scenarios were of similar order of magnitude as our standard method, except for the case of using the measured values for the spectroscopic correlation function parameters ($r_0$ and $\gamma$) in each z-bin instead of a fit. This yielded a bias in $\sigma_z$ of $\sim-0.009$. From this we see that the methods used to measure correlation parameters need to be considered carefully, since inferior methods can cause the bias to become comparable to random errors. 

From equation 13 in \cite{2008ApJ...684...88N}, the predicted errors in $\langle z \rangle$  using the weak clustering formalism are essentially identical to the errors in $\sigma_z$; that is true to $\sim 20$\% in our results.  This error is a function of $\sigma_z$, as well as the surface density of photometric objects on the sky, $\Sigma_p$, the number of objects per unit redshift of the spectroscopic sample, $dN_s/dz$, and the cross correlation parameters, $\gamma_{sp}$ and $r_{0,sp}$. We use the mean values of these parameters from our catalogs and find that the predicted error on both parameters is $\sigma=1.064\times10^{-3}$. This is considerably smaller than our measured error, which is not surprising given the extra error terms in the correlation function discussed in $\S\ref{sec:correrrors}$.

Our analysis throughout this paper has considered the case of a single-peaked, Gaussian selection function for placing objects in a photometric bin.  However, different distributions would yield similar results, as the error in the recovery of $\phi_p(z)$ at a given redshift depends primarily on the characteristics of the spectroscopic sample and the overall size of the photometric sample, but not $\phi_p(z)$ itself \citep{2008ApJ...684...88N}.  We illustrate this in Fig. \ref{fig:phi2peak}, where we have applied the same analysis techniques described above (and laid out in the recipe in \S\ref{sec:conclusion}) for a selection function that consists of two equal-amplitude Gaussian peaks centered at $z=0.5$ and $z=1.0$, each with $\sigma_z=0.1$; this figure can be compared to the right panel of Fig. \ref{fig:phicvar}.  We note that, since in this scenario the objects selected are less concentrated in redshift, the effects of bias evolution (as predicted by the semi-analytic models used) should be greater here than in our standard case, but our recovery remains accurate.

\section{Conclusion}
\label{sec:conclusion}
Section \ref{sec:method} has described in detail the steps we took to recover the redshift distribution, $\phi_p(z)$, of a photometric sample by cross\hyp{}correlating with a spectroscopic sample of known redshift distribution. We will now summarize the procedure used to make this calculation, to facilitate its application to actual data sets.

\begin{list}{\labelitemi}{\leftmargin=1em}
\item \textbf{Obtain the necessary information for each sample; RA, dec and redshift for the spectroscopic sample, and RA and dec for the photometric sample.}
\item \textbf{Create the random catalogs for each sample. (\S \ref{sec:autospec}-\ref{sec:crossphi})}
\item \textbf{Calculate the data-data, data-random, and random-random paircounts for each correlation function.}
\begin{list}{\labelitemi}{\leftmargin=1em}
\item For $w_p(r_p)$: bin the spectroscopic sample and its corresponding random catalog in redshift. In each spectroscopic z-bin, calculate $\Delta r_p$ and $\Delta \pi$ for each pair and bin the pair separations into a grid of $\log(r_p)$ and $\pi$. Then sum the paircounts in the $\pi$ direction. (\S \ref{sec:autospec})
\item For $w_{pp}(\theta)$: using the '$p$' sample and its random catalog, calculate $\Delta \theta$ for each pair and bin the pair separations into log($\theta$) bins. (\S \ref{sec:autophot})
\item For $w_{sp}(\theta,z)$: bin the spectroscopic sample and its corresponding random catalog in redshift. For each spectroscopic z-bin, calculate the pair separations, $\Delta \theta$, for pairs between the '$s$' and '$p$' samples and their random catalogs and bin them into $\log(\theta)$ bins. (\S \ref{sec:crossphi})
\end{list}

\item \textbf{Use the paircounts to calculate the correlation functions using standard estimators (e.g. Landy \& Szalay). (\S \ref{sec:autospec}-\ref{sec:crossphi})}

\item \textbf{Calculate the parameters of $w_p(r_p)\ (r_{0,ss}(z), \gamma_{ss}(z))$ and $w_{pp}(\theta)\ (A_{pp}, \gamma_{pp})$ by fitting as described above.  (\S \ref{sec:autospec}-\ref{sec:autophot})}

\item \textbf{Use the autocorrelation parameters along with an initial guess of $r_{0,pp}$ (e.g. $r_{0,pp}\sim r_{0,ss}$) to calculate $r^{\gamma_{sp}}_{0,sp}(z)=(r^{\gamma_{ss}}_{0,ss}r^{\gamma_{pp}}_{0,pp})^{1/2}$. (\S \ref{sec:autophot})} This gave a more accurate reconstruction of $\phi_p(z)$ (reducing $\chi^2$ by 33\%) than the assumption $r_{0,pp}=$ constant; in fact, a calculation of $\xi_{pp}(r)$ from the simulation sample directly showed $r_{0,pp}$ to have similar behavior to $r_{0,ss}$. Using a linear fit of $r_{0,ss}(z)$ and $\gamma_{ss}(z)$ reduced $\chi^2$ by $\sim32\%$ compared to utilizing the noisier reconstructed values in each z-bin.

\item \textbf{Estimate $\gamma_{sp}=(\gamma_{ss}+\gamma_{pp})/2$}. Using this $\gamma_{sp}$, calculate the amplitude, $A_{sp}(z)$, of $w_{sp}(\theta,z)$ by fitting as described above. (\S \ref{sec:crossphi}) We fit over the range $0.001^{\circ} <\theta< 0.1^{\circ}$. We found that fitting over this smaller $\theta$ range resulted in smaller errors in the amplitude, $A_{sp}(z)$, which reduced the error in $\phi_p(z)$ for each z-bin by $\sim25\%$ on average. We fix $\gamma_{sp}$ because of degeneracies between $\gamma_{sp}$ and $A_{sp}$ when fitting them simultaneously. This degeneracy is especially strong in regions where $\phi_p(z)$ is small. We also tried modeling $\gamma_{sp}$ as constant with z using the arithmetic mean of $\gamma_{ss}(z=0.77)$ and $\gamma_{pp}$; however, that method increased the $\chi^2$ of the final fit by $\sim20\%$.

\item \textbf{Combining the results of the last two steps and the assumed cosmology, calculate $\phi_p(z)$ using equation \ref{eq:phi}. (\S \ref{sec:crossphi})} We also tried calculating $\phi_p(z)$ using the integrated cross\hyp{}correlation function, $\tilde w(z)$, integrating to an angle equivalent to a comoving distance $r_{max}=10h^{-1}$ Mpc (Newman 2008); however, that method produced inferior results.

\item \textbf{Using $\phi_p(z)$, along with the calculated $A_{pp}$ and $\gamma_{pp}$, in equation \ref{eq:wpp} gives a new $r_{0,pp}$, which is then used to recalculate $r^{\gamma_{sp}}_{0,sp}(z)$. Putting this back into equation \ref{eq:phi} gives a new $\phi_p(z)$. This is repeated until convergence is reached. (\S \ref{sec:crossphi})}

\item \textbf{To recover the underlying/universal distribution of objects of the type selected for the photometric sample, rather than the distribution within the specific fields chosen for observation, correct for sample/cosmic variance using the fluctuations in the redshift distribution of the spectroscopic; i.e., construct a smooth function describing the overall redshift distribution of the spectroscopic sample, $\langle dN_s/dz\rangle$, and divide $\phi_p(z)$ by the ratio $dN_s/dz/\langle dN_s/dz\rangle$. (\S \ref{sec:errorest})}

\end{list}

We have shown in this paper that by exploiting the clustering of galaxies at similar redshifts we can accurately recover the redshift distribution of a photometric sample using its angular cross\hyp{}correlation with a spectroscopic sample of known redshift distribution, using mock catalogs designed to match the DEEP2 Galaxy Redshift Survey. This test includes the impact of realistic bias evolution and cosmic variance. Our error estimates for the recovered mean and standard deviation of the distribution are larger than those predicted previously, but improvements could be obtained either by using more optimal correlation function estimators or by surveying the same number of galaxies distributed over a wider area of sky.  Based on these tests we expect that this technique should be able to deliver the performance needed for dark energy experiments.

In a recent paper \citep{2009arXiv0910.3683S}, cross\hyp{}correlation techniques were applied to mock data generated by populating a single time slice of an N-body dark matter simulation using various halo models. They develop a pipeline for calculating the redshift distribution of a photometric sample using cross\hyp{}correlation measurements and the autocorrelation of a spectroscopic sample, $\xi_{ss}(r,z)$. They do not attempt to model the bias although they do examine how varying the bias of the two samples affects the reconstruction (i.e. using radically different halo models). The catalogs constructed to test their method are significantly larger in volume than our individual mock catalogs, and while the number of objects in their photometric sample is comparable to ours, their spectroscopic sample is much smaller, which would be expected to lead to larger errors (Newman 2008), as observed. Another major difference is the use of a smoothness prior in reconstruction, which was not done here. While \citet{2009arXiv0910.3683S} found that cross\hyp{}correlation techniques were generally successful in reconstructing redshift distributions, these conclusions were primarily qualitative due to the limited sample sizes and source densities of the mock samples used, along with less-optimal correlation measurement techniques. In this paper, we have used simulations which include much less massive halos, allowing us to perform quantitative tests of cross\hyp{}correlation techniques using sample sizes and source densities comparable to those which will be used in realistic applications.

Several techniques for calibrating photometric redshifts using only photometric data have also been developed \citep{2006ApJ...651...14S, 2009arXiv0910.4181Z, 2010arXiv1002.2266B, 2009arXiv0910.2704Q}; in general, such techniques require priors or assumptions on biasing which can be relaxed or tested in spectroscopic cross\hyp{}correlation measurements. In \cite{2009arXiv0910.2704Q}, spectroscopic/photometric cross\hyp{}correlation techniques have now been applied to real data using the COSMOS dataset. Using data from a single field, they are able to determine typical photo-z uncertainties well, even when ignoring the effects of bias evolution. However, when constraining catastrophic photo-z errors, methods which ignore these effects should break down, as bias evolution should be a much greater problem over broad redshift intervals than in the core of the photo-z error distribution.

In future work, we will explore alternate methods of measuring correlation functions that are invariant to the variance in the integral constraint (e.g. \cite{2007MNRAS.376.1702P}). This should reduce errors in the measurement of the redshift distribution, which we found to be larger than expected due to extra variance terms in the correlation function measurements not considered previously. We also plan to test this technique with mock catalogs in which photometric redshifts have been 'measured' on simulated LSST photometry, rather than simply assuming a redshift distribution. We will also apply this method to real data using photometric and spectroscopic samples from the AEGIS survey \citep{2007ApJ...660L...1D}.

The authors wish particularly to thank Darren Croton for developing the mock catalogs used and making them publicly available. We would also like to thank Andrew Hearin, Arthur Kosowsky, David Wittman, Michael Wood-Vasey, Andrew Zentner, Tony Tyson, Gary Bernstein, Nikhil Padmanabhan, Dragan Huterer, Hu Zhan, and in particular Alexia Schulz for useful discussions during the course of this work, and also Ben Brown and Brian Cherinka for their technical expertise in performing these calculations. This work has been supported by the United States Department of Energy.

\bibliography{adsreferences}
\clearpage

\end{document}